\documentclass[journal]{IEEEtran}
\usepackage{amsmath, amssymb, cite, tikz, graphicx, xspace, mathrsfs, mathtools, psfrag, chemarrow, subfigure,color,soul,graphicx,mathtools,adjustbox}
\usepackage{kbordermatrix}
\usepackage{enumitem}
\usepackage{bbm}
\usepackage[utf8]{inputenc}
\usepackage{booktabs}
\usepackage{amscd}
\usepackage{amsmath}
\usepackage{amssymb}
\usepackage{amsthm}

\usepackage{epsfig}
\usepackage{verbatim}
\usepackage{graphicx}
\usepackage{amsthm}
\usepackage{color}
\usepackage{ multirow }
\usepackage[noend]{algpseudocode}
\usetikzlibrary{patterns}
\usetikzlibrary{angles} 
\usepackage{eucal}
\usepackage{algorithm}
\usepackage{tikz}
\usetikzlibrary{shapes, arrows, decorations.markings, arrows.meta}
\usetikzlibrary{patterns} 
\usetikzlibrary{arrows,automata,positioning,chains,calc}
\usetikzlibrary{quotes,angles}
\newtheorem{remark}{Remark}
\makeatletter
\def\BState{\State\hskip-\ALG@thistlm}
\makeatother
\newtheorem{theorem}{Theorem}
\newtheorem{lemma}[theorem]{Lemma}
\begin{document}
	\setlength{\abovecaptionskip}{-3pt}
	\setlength{\belowcaptionskip}{1pt}
	\setlength{\floatsep}{1ex}
	\setlength{\textfloatsep}{1ex}	
	\title{{Age of Information Versions: a Semantic View of Markov Source Monitoring } }
	
	\author{Mehrdad Salimnejad, Marios Kountouris, \IEEEmembership{Fellow, IEEE}, Anthony Ephremides  \IEEEmembership{Life Fellow, IEEE}\\ and Nikolaos Pappas, 
		\IEEEmembership{Senior Member, IEEE}.
		\thanks{M. Salimnejad and N. Pappas are with the Department of Computer and Information Science Linköping University, Sweden, email: \{\texttt{mehrdad.salimnejad, nikolaos.pappas\}@liu.se}. M. Kountouris is with the Department of Computer Science and Artificial Intelligence, Andalusian Research Institute in Data Science and Computational Intelligence (DaSCI), University of Granada, Spain, email: \texttt{mariosk@ugr.es}. A. Ephremides is with the Electrical and Computer Engineering, University of Maryland, College Park, MD, USA, email: \texttt{etony@umd.edu}\\ The work of M. Salimnejad and N. Pappas has been supported in part by the Swedish Research Council (VR), ELLIIT, Zenith, and the European Union (ETHER, 101096526). The work of M. Kountouris has received funding from the European Research Council (ERC) under the European Union’s Horizon 2020 research and innovation programme (Grant agreement No. 101003431).}}
	
	\maketitle
	\begin{abstract}
		We consider the problem of real-time remote monitoring of a two-state Markov process, where a sensor observes the state of the source and makes a decision on whether to transmit the status updates over an unreliable channel or not. We introduce a modified randomized stationary sampling and transmission policy where the decision to perform sampling occurs probabilistically depending on the current state of the source and whether the system was in a sync state during the previous time slot or not. We then propose two new performance metrics, coined \emph{the Version Innovation Age (VIA)} and \emph{the Age of Incorrect Version (AoIV)} and analyze their performance under the modified randomized stationary and other state-of-the-art sampling and transmission policies. Specifically, we derive closed-form expressions for the distribution and the average of VIA, AoIV, and Age of Incorrect Information (AoII) under these policies. Furthermore, we formulate and solve three constrained optimization problems. The first optimization problem aims to minimize the average VIA subject to constraints on the time-averaged sampling cost and time-averaged reconstruction error. In the second and third problems, the objective is to minimize the average AoIV and AoII, respectively, while considering a constraint on the time-averaged sampling cost. Finally, we compare the performance of various sampling and transmission policies and identify the conditions under which each policy outperforms the others in optimizing the proposed metrics.
	\end{abstract}
	
	\section{Introduction}
	Timely delivery of relevant status update packets from an information source has become increasingly crucial in various real-time communication systems, in which digital components remotely monitor and control physical entities \cite{abd2019role, shreedhar2019age}. These systems require reliable and timely exchange of useful information, coupled with efficient distributed processing, to facilitate optimal decision-making in applications such as industrial automation, collaborative robotics, and autonomous transportation systems. These challenging requirements gave rise to \emph{goal-oriented semantics-empowered communications} \cite{kountouris2021semantics,tolga21SP,popovski2020semantic,PetarProc2022, GunduzJSAC23}, a novel paradigm that considers the usefulness, the timeliness \cite{kaul2012Real,stamatakis2019control,pappas2022agebook,stamatakis2024semantics} and the innate and contextual importance of information to generate, transmit, and utilize data in time-sensitive and data-intensive communication systems. 
	Recently, a new metric called Version Age of Information (VAoI) has been introduced in \cite{yates2021age}, where each update at the source is considered as a new version, thus quantifying how many versions out-of-date the information on the monitor is, compared to the version at the source. Several studies have considered the VAoI as a key performance metric of timeliness in information in networks \cite{yates2021age,yates2021timely,buyukatesversion,kaswan2022timely,kaswan2022susceptibility,kaswan2022age,mitra2023age,abd2023distribution,KaswanTCOM2023,KaswanJSAC2023,bastopcu2023role,mitra2023learning,delfani2023version,karevvanavar2023version}. The scaling of the average VAoI in gossip networks of different sizes and topologies is investigated in \cite{yates2021age,yates2021timely,buyukatesversion,kaswan2022timely,kaswan2022susceptibility,kaswan2022age,mitra2023age,abd2023distribution,KaswanTCOM2023,KaswanJSAC2023,bastopcu2023role}. A learning-based approach to minimize the overall average VAoI of the worst-performing node in sparse gossip networks is employed in \cite{mitra2023learning}.  
	The authors in \cite{delfani2023version} studied the problem of minimizing the average VAoI using a Markov Decision Process (MDP) in a scenario where an energy harvesting (EH) sensor updates the gossip network via an aggregator equipped with caching capabilities. 
	The work \cite{karevvanavar2023version} considered the problem of minimizing VAoI over a fading broadcast channel, employing a non-orthogonal multiple access (NOMA) scheme.
	
	Prior works on VAoI rely on the occurrence of \emph{content change} at the source, while the destination nodes only demand the latest version of the information from the source, regardless of the \emph{content} of the information. In other words, VAoI exclusively focuses on changes occurring in the source's content, disregarding the significance and the utility of that information. Several semantics-aware metrics \cite{maatouk2020age,pappas2021goal,jayanth23,MSalimnejadTCOM2024,MSalimnejadJCN2023,fountoulakis2023goal,luo2024semantic,cocco2023remote} have made a step in that direction, without though investigating the evolution of versions. In this work, we propose two new semantics-aware metrics, coined \emph{Version Innovation Age} (VIA), and \emph{ Age of Incorrect Version} (AoIV), which take into account both the content and the version evolution of the information source.
	\par In this paper, we examine a time-slotted communication system where a sampler makes decisions to perform sampling of a two-state Markov process, acting as the information source. After deciding to sample, the transmitter sends the sample in packet form to a remote receiver over an unreliable wireless channel. Then, at the receiver, real-time reconstruction of the information source is conducted based on the successfully received samples. It is assumed that the system is in a sync state if the state of the source matches the state of the reconstructed source; otherwise, the system is considered to be in an erroneous state. Furthermore, a specific action is executed at the receiver based on the estimated state of the information source. 
	We introduce a new sampling and transmission policy, termed \emph{modified randomized stationary} policy. To evaluate the system performance, we derive general expressions for the distribution and average of the VIA and AoIV, as well as for the Age of Incorrect information (AoII) \cite{maatouk2020age}, under the modified randomized stationary policy and other state-of-the-art sampling and transmission policies introduced in \cite{pappas2021goal, MSalimnejadTCOM2024}. In addition, we formulate and solve three constrained optimization problems. In the first optimization problem, the aim is to minimize the average VIA with constraints on the time-averaged sampling cost and time-averaged reconstruction error metric as proposed in \cite{MSalimnejadTCOM2024}. In the second and third problems, we minimize the average AoIV and AoII, respectively, under time-averaged sampling cost as a constraint. Furthermore, the impact of various system parameters on optimizing the proposed metrics is investigated through analytical and numerical results.
	\vspace{-0.4cm}
	\subsection{Related Works}
	\par The primary focus of this study is to design joint source sampling, transmission, and reconstruction policies, considering the dynamics of the information source, to enable real-time remote tracking of Markovian sources for actuation purposes. Several studies have considered the problem of real-time remote monitoring of Markovian sources\cite{nayyar2013optimal,chakravorty2014optimal,shi2012scheduling,wu2018optimal,chakravorty2015distortion, chakravorty2016fundamental,chakravorty2019remote}. In \cite{nayyar2013optimal}, the authors investigated the problem of determining optimal communication and estimation policies for a remote estimation scenario involving an energy harvesting source that observes a finite state Markov process. In \cite{chakravorty2014optimal}, optimal transmission and estimation strategies are derived for a noiseless communication system in which a sensor observes a first-order Markov process. In \cite{shi2012scheduling}, an optimal transmission policy has been proposed for two Gauss-Markov systems where the states are measured by two sensors. The results of this study have been extended in \cite{wu2018optimal}, where the authors examine a scenario involving multiple sensors and processes. The problem of optimal real-time transmission of Markov processes under communication constraints is presented in \cite{chakravorty2015distortion, chakravorty2016fundamental, chakravorty2019remote}. The main objective of these studies is to propose sampling and transmission strategies aimed at minimizing estimation errors, without considering the significance and usefulness of the information source or its impact on actuation. In \cite{maatouk2020age,pappas2021goal,jayanth23,MSalimnejadTCOM2024,MSalimnejadJCN2023,fountoulakis2023goal,luo2024semantic,cocco2023remote}, the authors proposed several semantics-aware metrics that capture the effectiveness of information sources by leveraging synergies between data processing, information transmission, and signal reconstruction. In \cite{maatouk2020age}, the authors proposed the AoII metric, which measures the elapsed time between the present and the most recent time when the content of the sample at its source matched the content of the sample stored at the receiver. A new semantics-aware metric named the \emph{cost of actuation error} was introduced in \cite{pappas2021goal}, which captures the significance of erroneous actions at the receiver side. The work \cite{jayanth23} considered a real-time monitoring system where a source monitors a two-state Markov process, and a receiver acquires timely and accurate state information. The authors derived an optimal transmission policy aimed at minimizing the average expected distortion while considering constraints on the average expected AoI and source costs. In \cite{MSalimnejadTCOM2024,MSalimnejadJCN2023}, the authors introduced new goal-oriented joint sampling and transmission policies for real-time tracking and source reconstruction Markov processes. They also proposed new performance metrics, namely \emph{consecutive error} and \emph{cost of memory error} \cite{MSalimnejadTCOM2024}, and \emph{importance-aware consecutive error} \cite{MSalimnejadJCN2023}. 
	The work \cite{fountoulakis2023goal} proposed an optimal transmission policy to minimize the cost of actuation error in a single-source constrained system, which was extended to multiple sources and processes in \cite{luo2024semantic}. \cite{cocco2023remote} studied the remote monitoring problem in a system where two-state Markov sources send status updates to a common receiver over a slotted ALOHA random access channel. Then, the performance of the system is analyzed in terms of state estimation entropy, which measures the uncertainty at the receiver regarding the sources' state.
	\vspace{-0.4cm}
	\subsection{Contributions}
	\par The key contributions of this paper are summarized below:
	\begin{enumerate}
		\item We propose a new sampling and transmission policy, coined \emph{modified randomized stationary}, which incorporates varying sampling probabilities based on the current state of the source and whether the system was in a sync state or not. This metric is relevant in scenarios where the power budget for sampling and transmission actions is limited, and the decision to perform sampling depending on the sync state of the system in the previous state could have a varying impact on system performance. In such scenarios, it is crucial to adopt different sampling frequencies.
		\item We introduce two new semantics-aware metrics, namely \emph{VIA} and \emph{AoIV}, which consider both the content and the version evolution of the information source. To assess the performance of the system, we obtain the distribution and average of the proposed metrics and AoII under the modified randomized stationary sampling policy, as well as the randomized stationary, change-aware, and semantics-aware policies introduced in \cite{pappas2021goal, MSalimnejadTCOM2024}.
		\item We formulate and solve three constrained optimization problems. In the first optimization problem, the objective is to minimize the average VIA subject to constraints on the time-averaged sampling cost and time-averaged reconstruction error. In the second and third optimization problems, our objective is to minimize the average AoIV and AoII, respectively, while considering a constraint on the time-averaged sampling cost. By solving these optimization problems, we identify the conditions under which each policy outperforms the others in optimizing the proposed metrics.
	\end{enumerate}
	
	\vspace{-0.4cm}
	\section{System Model}
	\label{system_model}
	\par
\begin{figure}[!t]
		\centering
		\footnotesize
		\begin{tikzpicture}[start chain=going left,>=stealth,node distance=2cm,on grid,auto]
			\footnotesize
			\node[state, on chain]               at (0,-2)  (2) {$1$};
			\node[state, on chain]              at  (0,-2)   (1) {$0$};
			\path[->]
			(1)   edge[loop above] node  {$1-p$}   (1)
			(1) edge  [bend left=30] node {$p$} (2)
			
			(2) edge  [bend left=30] node {$q$} (1)
			(2) edge  [loop above] node {$1-q$} (2)
			;
			\draw [line width=0.35mm](-2.5,0)--(0.5,0)--(0.5,-2.8)--(-2.5,-2.8)--(-2.5,0);
			\node(a) at (-1,-0.4)  {\normalsize{{Source}},\! $X(t)$};
			\draw [line width=0.35mm](0.5,-1.4) -- (1,-1.4);
			\draw [line width=0.35mm](1,-1.)to(1.7,-1.4);
			\draw [line width=0.35mm](1.69,-1.4)to(2.1,-1.4);
			
			\draw[line width=0.3mm,-{Stealth[length=2mm]}] (1.61,-1.) arc (90:205:5mm) ;
			\draw [line width=0.35mm](2.53,-1.4) circle (0.45cm);
			\node(b) at (2.53,-1.4)  {\small	$\text{Tx}$};
			\draw [-{Stealth[length=3mm, width=2mm]},dashed](3,-1.4) -- (4.29,-1.4);
			\draw [line width=0.35mm](4.73,-1.4) circle (0.45cm);
			\node(c) at (4.73,-1.4)  {\small	$\text{Rx}$};
			
			\node(e) at (1.2,-0.8)  {\footnotesize$\text{ Sampler}$};
			\draw [-{Stealth[length=3mm, width=2mm]}](4.73,-.95) -- (4.73,-0.21);
			\filldraw [line width=2mm](4.28,-.12) -- (5.18,-.12);
			\node(d) at (5.15,-.6)  {\small$\hat{X}(t)$};

			\draw [line width=0.25mm](4.61,-.1) -- (4.37,0.35);
			
			\draw [line width=0.25mm](4.89,-.1) -- (4.61,0.45);
			
			\draw [line width=0.35mm](4.45,0.5) circle (0.17cm);
			\filldraw [line width=0.35mm](4.45,0.5) circle (0.06cm);
			
			\draw [line width=0.25mm](4.45,0.68) -- (5.15,0.9);
			\draw [line width=0.25mm](4.61,0.43) -- (5.2,0.73);
			
			\draw [line width=0.9mm](5.23,0.7) -- (5.12,0.935);
			
			\draw [line width=0.2mm](5.15,0.9) -- (5.28,1.07);
			\draw [line width=0.2mm](5.28,1.07) -- (5.47,1.04);
			
			\draw [line width=0.2mm](5.23,0.76) -- (5.42,0.75);
			\draw [line width=0.2mm](5.42,0.75) -- (5.52,0.91);
		\end{tikzpicture}
		\vspace*{2ex}
		\caption{Real-time monitoring of a Markovian source over a wireless channel.}
		\label{system_model_fig}
	\end{figure}
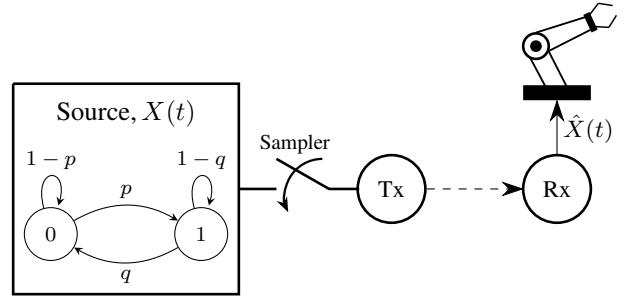

 We consider a time-slotted communication system in which a sampler conducts sampling of an information source, denoted as $X(t)$, at time slot $t$, as shown in Fig. \ref{system_model_fig}. The transmitter then forwards the sampled information in packets over a wireless communication channel to the receiver. The information source is modeled as a two-state discrete-time Markov chain (DTMC) $\{X(t), t \in \mathbb{N}\}$. Therein, the state transition probability $\mathrm{Pr}\big[X(t+1)=j \big|X(t)=i\big]$ represents the probability of transitioning from state $i$ to $j$ and can be defined as $\mathrm{Pr}\big[X(t+1)=j\big|X(t)=i\big] = \mathbbm{1} (i=0,j=0)(1-p)+\mathbbm{1}(i=0,j=1)p+\mathbbm{1}(i=1,j=0)q+\mathbbm{1}(i=1,j=1)(1-q)$, where $\mathbbm {1}(\cdot)$ is the indicator function. We denote the action of sampling at time slot $t$ by $\alpha^{\text{s}}(t)=\{0,1\}$, where $\alpha^{\text{s}}(t)=1$ if the source is sampled and $\alpha^{\text{s}}(t)=0$ otherwise\footnote{We assume that when sampling occurs at time slot $t$, the transmitter sends the sample immediately at that time slot.}. At time slot $t$, the receiver constructs an estimate of the process $X(t)$, denoted by $\hat{X}(t)$, based on the successfully received samples. For the wireless channel between the source and the receiver, we assume that the channel state $h(t)$ equals $1$ if the information source is sampled and successfully decoded by the receiver and $0$ otherwise. We define the success probability as $p_{s}=\mathrm{Pr}[h(t)=1]$. At time slot $t$, when the source is sampled, we assume that with probability $p_{s}$, the system is in a sync state, i.e., $X(t)=\hat{X}(t)$. Otherwise, if the system is in an erroneous state, the state of the reconstructed source remains unchanged, i.e., $\hat{X}(t)=\hat{X}(t-1)$. Acknowledgment (ACK)/negative-ACK(NACK) packets are used to inform the transmitter about the success or failure of transmissions. ACK/NACK packets are assumed to be delivered to the transmitter instantly and without errors\footnote{An ACK/NACK feedback channel is required for the modified randomized stationary and semantics-aware policies.}. Therefore, the transmitter has accurate information about the reconstructed source state at time slot $t$, i.e., $\hat{X}(t)$. In addition, we assume that the corresponding sample is discarded in the event of a transmission failure (packet drop channel). Here, we propose a new sampling and transmission policy, named the \emph{modified randomized stationary} policy, where the generation of a new sampling is triggered probabilistically. Under this policy, no source sampling occurs if the state of the source at the current time slot matches the state of the reconstructed source at the previous time slot. Otherwise, we employ different sampling probabilities based on whether the system was in a sync state during the previous time slot or not. Furthermore, for comparison purposes, we adopt three other relevant sampling policies, namely the \emph{randomized stationary}, \emph{change-aware}, and \emph{semantics-aware} policies proposed in \cite{pappas2021goal, MSalimnejadTCOM2024}. Below, we briefly describe these policies.
	\begin{enumerate}
		\item  \emph{Randomized Stationary (RS)}: a new sample is generated probabilistically at each time slot without considering whether the system is in sync or not. We assume that the probability of sampling at time slot $t$ is $p_{\alpha^{\text{s}}}\!\!=\!\!\mathrm{Pr}[\alpha^{\text{s}}(t)\!=\!1]$. We also define $\mathrm{Pr}[\alpha^{\text{s}}(t)\!=\!0]\!=\!1\!-\!p_{\alpha^{\text{s}}}$ as the probability that the source is not sampled at time slot $t$.
		\item \emph{Modified Randomized Stationary (MRS)}: in this policy, at time slot $t$, the source is not sampled if the state of the source matches the state of the reconstructed source at the previous time slot, i.e., $X(t) = \hat{X}(t-1)$. This condition indicates that the system is in a sync state and thus sampling is unnecessary. Otherwise, the generation of a new sample at time slot $t$ is triggered probabilistically. Here, we assume that when the system was in a sync state at time slot $t-1$ but the state of the source at time slot $t$ differs from that at time slot $t-1$, the sampler performs sampling with probability $q_{\alpha^{\text{s}}_{1}}$, where $q_{\alpha^{\text{s}}_{1}} = \mathrm{Pr}[\alpha^{\text{s}}(t) = 1 \mid X(t) \neq {X}(t-1), X(t-1) = \hat{X}(t-1)]$. Furthermore, when the state of the source at time slot $t-1$ and $t$ differs from the state of the reconstructed source at time slot $t-1$, the sampler performs sampling with probability $q_{\alpha^{\text{s}}_{2}}$, where $q_{\alpha^{\text{s}}_{2}} = \mathrm{Pr}[\alpha^{\text{s}}(t) = 1 \mid X(t) \neq \hat{X}(t-1), X(t-1) \neq \hat{X}(t-1)]$.
		\item \emph{Change-aware}: a new sample is conducted at time slot $t$ if the state of the source differs from that at time slot $t-1$, i.e., $X(t)\neq X(t-1) $, regardless of whether the system is sync or not.
		\item \emph{Semantics-aware}: the sampler performs sampling at time slot $t$ when the state of the source at time slot $t$ is not equal to the state of the reconstructed source at the previous time slot, i.e., $X(t) \neq \hat{X}(t-1)$.
	\end{enumerate}
	\vspace{-0.1cm}
	\section{Performance Metrics and Analysis}
	In this section, we study the effect of the \emph{semantics of information} at the receiver. For that, we propose and analyze two new performance metrics, termed VIA and AoIV, which jointly quantify \emph{both the timing and importance aspects of information}. Furthermore, we analyze the performance of the average AoII.
	\vspace{-0.4cm}
	\subsection{Version Innovation Age (VIA)}
	\par Before introducing the first new metric, we review the VAoI proposed in \cite{yates2021age}. We assume that each update at the information source represents a version. At time slot $t$, $V_{\text{S}}(t)$ represents the version at the source, while $V_{\text{R}}(t)$ represents the version at the receiver. The VAoI at the receiver is then defined as $\text{VAoI}(t) = V_{\text{S}}(t)-V_{\text{R}}(t)$. This metric relies on changes in the information source, but the content of the information source is not considered important. Therefore, we introduce a new metric named VIA, which measures the number of outdated versions at the receiver compared to the source when the source is in a specific state. In other words, the VIA differs when the source is in state $i$ compared to when it is in state $j$. Since the VAoI solely considers changes in the source state and does not depend on the specific state of the source, this metric cannot be defined for a specific state. We define the evolution of VIA as follows
	\begin{align}
		\label{Version_AoI}
		{\text{VIA}}\big(t\big)\!\!=\!\!
		\begin{cases}
			\!\!{\text{VIA}}(t-1), &\!\!\!\!\parbox[t]{6cm}{{$X(t)\!=\!X(t-1)$ \!\text{and} $\{\alpha^{\text{s}}(t)\!=\!0,\\ \text{or}\hspace{0.1cm}(\alpha^{\text{s}}(t)\!=\!1, h(t)\!=\!0)\}$,}\\}\\
			\!\!{\text{VIA}}(t-1)\!+\!1, & \!\!\!\!\parbox[t]{9cm}{{$X(t)\!\neq\! X(t-1)$ \!\text{and} $\{\alpha^{\text{s}}(t)\!=\!0,\\ \text{or} \hspace{0.1cm} (\alpha^{\text{s}}(t)\!=\!1, h(t)\!=\!0)\}$,}\\}\\
			0, &\!\!\! \parbox[t]{9cm}{$\alpha^{\text{s}}(t)\!=\!1, h(t)=1$.}\\
		\end{cases}
	\end{align}
	Using \eqref{Version_AoI}, for a two-state DTMC information source depicted in Fig. \ref{system_model_fig}, and applying the total probability theorem, we can express the probability that the VIA at time slot $t$ equals $i \geqslant 0$ as follows
	\begin{align}
		\label{VIA_Pr}
		\!\!\!\!\!\!\mathrm{Pr}\big[\text{VIA}(t)=i\big]&\!\!=\! \mathrm{Pr}\big[X(t)=0,\text{VIA}(t)=i\big]\notag\\
		&\!\!+\!\mathrm{Pr}\big[X(t)=1,\text{VIA}(t)=i\big]
		=\pi_{0,i}+\pi_{1,i}.
	\end{align}
	Note that $\pi_{0,i}$ and $\pi_{1,i}$ in \eqref{VIA_Pr} are the probabilities obtained from the stationary distribution of the two-dimensional DTMC describing the joint status of the original source regarding the current state of the VIA, i.e., $\big(X(t), {\text{VIA}}(t)\big)$.
	
	\begin{lemma}
		\label{theorem_VAoI}
		For a two-state DTMC information source, the stationary distribution $\pi_{j,i}$ for the RS policy is given by\footnote{Since, at each time slot, the state of the VIA does not depend on the state of the reconstructed source, we only investigate the performance of this metric for RS and change-aware policies.}
		\begin{align}
			\!\!\pi_{0,i}\!\!&=\!\!\frac{p^kq^wp_{\alpha^{\text{s}}}p_{s}\big(1\!-\!p_{\alpha^{\text{s}}}p_{s}\big)^i}{(p\!+\!q)\big(p\!+\!(1\!-\!p)p_{\alpha^{\text{s}}}p_{s}\big)^w\big(q\!+\!(1\!-\!q)p_{\alpha^{\text{s}}}p_{s}\big)^k},
			i\!=\!0,1,\ldots.\notag\\
			\!\!\pi_{1,i}\!\!&=\!\!\frac{p^wq^kp_{\alpha^{\text{s}}}p_{s}\big(1\!-\!p_{\alpha^{\text{s}}}p_{s}\big)^i}{(p\!+\!q)\big(p\!+\!(1\!-\!p)p_{\alpha^{\text{s}}}p_{s}\big)^k\big(q\!+\!(1\!-\!q)p_{\alpha^{\text{s}}}p_{s}\big)^w}, i\!=\!0,1,\ldots.\label{pi0i_pi1i_AoI_RS}
		\end{align}
		where $k$ and $w$ are given by
		\begin{align}
			\label{kw_AoI}
			k=
			\begin{cases}
				\frac{i}{2}, &\mod\{i,2\}=0,\\
				\frac{i+1}{2}, &\mod\{i,2\}\neq 0.
			\end{cases}\notag\\
			\hspace{0.2cm}w=
			\begin{cases}
				\frac{i+2}{2}, &\mod\{i,2\}=0,\\
				\frac{i+1}{2}, &\mod\{i,2\}\neq 0.
			\end{cases}
		\end{align}
	\end{lemma}
	
	\begin{IEEEproof} 
		See Appendix \ref{Appendix_LemmaPij_VAoI}.
	\end{IEEEproof}
	
	Using \eqref{pi0i_pi1i_AoI_RS} and \eqref{kw_AoI}, we calculate the average VIA, $\overline{\text{VIA}}$, as\footnote{The convergence condition for \eqref{Avg_AoIV_RS} is $\!\!\frac{\sqrt{pq}\big(1\!-\!p_{\alpha^{\text{s}}}p_{s}\big)}{\sqrt{\!\big(p\!+\!(1-p)p_{\alpha^{\text{s}}}p_{s}\big)\!\!\big(q\!+\!(1-q)p_{\alpha^{\text{s}}}p_{s}\big)}}\!\!<\!\!1$.}
	\begin{align}
		\label{Avg_AoIV_RS}
		\overline{\text{VIA}} &= \sum_{i=0}^{\infty} i\mathrm{Pr}\big[{\text{VIA}}(t)=i\big] =  \sum_{i=1}^{\infty} i\big(\pi_{0,i}+\pi_{1,i}\big)\notag\\
		&=\frac{2pq\big(1-p_{\alpha^{\text{s}}}p_{s}\big)}{(p+q)p_{\alpha^{\text{s}}}p_{s}}.
	\end{align}
	
	\begin{remark}
		\label{remark_RS_CA_compare_AoIV}
		Using \eqref{Avg_AoIV_RS} and \eqref{Avg_AoIV_CA}, we can prove that for $0<p_{\text{s}}\leqslant1$, the RS policy has a lower average VIA compared to the change-aware policy if $\frac{2pq}{p+q+\big(2pq-p-q\big)p_{\text{s}}}\leqslant p_{\alpha^{\text{s}}}\leqslant1$. Otherwise, the change-aware policy outperforms the RS policy in terms of average VIA. 
	\end{remark}
	\begin{remark}
		\label{remark_RS_CA_AoIV_PE}
		Using (39) in \cite{MSalimnejadTCOM2024}, we can express $\overline{\text{VIA}}$ given in \eqref{Avg_AoIV_RS} and \eqref{Avg_AoIV_CA} as a function of the time-averaged reconstruction error, which is the probability that the system is in an erroneous state. For the RS policy, \eqref{Avg_AoIV_RS} is
		\begin{align}
			\label{PE_AoIV_RS}
			\overline{\text{VIA}}(P_{E})=\frac{\big[p+q+(1-p-q)p_{\alpha^{\text{s}}}p_{\text{s}}\big]P_{E}}{p_{\alpha^{\text{s}}}p_{\text{s}}}, 
		\end{align}
		where $P_{E}$ represents the time-averaged reconstruction error and in \eqref{PE_AoIV_RS} is given by
		\begin{align}
			\label{PE_RS}
			P_{E} = \frac{2pq(1-p_{\alpha^{\text{s}}}p_{\text{s}})}{(p+q)\big[p+q+(1-p-q)p_{\alpha^{\text{s}}}p_{\text{s}}\big]}.
		\end{align}
		Furthermore, for the change-aware policy, the expression given in \eqref{Avg_AoIV_CA} can be written as 
		\begin{align}
			\label{PE_AoIV_CA}
			\overline{\text{VIA}}(P_{E}) = \bigg(\frac{2}{p_{\text{s}}}-1\bigg)P_{E},
		\end{align}
		where $P_{E}$ in \eqref{PE_AoIV_CA} is calculated as
		\begin{align}
			\label{PE_CA}
			P_{E}=\frac{1-p_{\text{s}}}{2-p_{\text{s}}}.
		\end{align}
	\end{remark}
	
	\subsection{Age of Incorrect Version (AoIV)}
	\par A major issue with VIA is that when the state of the source changes and transmission fails, the VIA increases by one. However, the system may be in a sync state, i.e., $X(t) = \hat{X}(t)$. In other words, the VIA can still increase even if the system has perfect knowledge of the source's state. For that, we introduce another metric, termed AoIV, defined as the number of outdated versions at the receiver compared to the source when the system is in an erroneous state, i.e., $X(t) \neq \hat{X}(t)$. The evolution of the AoIV is
	\begin{align}
		\label{Version_AoII}
		{\text{AoIV}}\big(t\big) =
		\begin{cases}
			{\text{AoIV}}(t-1),&\parbox[t]{9cm}{{$X(t)=X(t-1),\\X(t)\neq\hat{X}(t)$}\\}\\ 
			{\text{AoIV}}(t-1)+1,&\parbox[t]{9cm}{{$X(t)\neq X(t-1),\\X(t)\neq\hat{X}(t)$}\\}\\ 
			0, &X(t)=\hat{X}(t).
		\end{cases}
	\end{align}
	Using \eqref{Version_AoII}, for a two-state DTMC information source as in Fig. \ref{system_model_fig}, and applying the total probability theorem, $\mathrm{Pr}\big[{\text{AoIV}}(t)\neq 0\big]$ is calculated as\footnote{For a two-state DTMC information source, the maximum value of AoIV is $1$.}
	\begin{align}
		\label{PrDAoIVI}
		\mathrm{Pr}\big[{\text{AoIV}}(t)\neq 0\big] &= \mathrm{Pr}\big[X(t)=0,\hat{X}(t)=1,{\text{AoIV}}(t)=1\big]\notag\\&+\mathrm{Pr}\big[X(t)=1,\hat{X}(t)=0,{\text{AoIV}}(t)=1\big]\notag\\&=\pi_{0,1,1}+\pi_{1,0,1},
	\end{align}
	where $\pi_{0,1,1}$ and $\pi_{1,0,1}$ are the probabilities obtained from the stationary distribution of the three-dimensional DTMC describing the joint status of the original and reconstructed source regarding the current state of the AoIV, i.e., $\big(X(t),\hat{X}(t),{\text{AoIV}}(t)\big)$. 
	\begin{lemma}
		\label{piijk_VAoII_RS}
		For a two-state DTMC information source, the stationary distribution $\pi_{i,j,k}, \forall i,j,k \in\{0,1\}$ for the RS  policy  is given by
		\begin{align}
			\label{VAoII_pi_ijk_RS}
			\pi_{0,0,0}
			&=\frac{q\big[q+(1-q)p_{\alpha^{\text{s}}}p_{s}\big]}{(p+q)\big[p+q+(1-p-q)p_{\alpha^{\text{s}}}p_{s}\big]}, \notag\\
			\pi_{0,1,1}
			&=\frac{pq\big(1-p_{\alpha^{\text{s}}}p_{s}\big)}{(p+q)\big[p+q+(1-p-q)p_{\alpha^{\text{s}}}p_{s}\big]},\notag\\
			\pi_{1,1,0}
			&=\frac{p\big[p+(1-p)p_{\alpha^{\text{s}}}p_{s}\big]}{(p+q)\big[p+q+(1-p-q)p_{\alpha^{\text{s}}}p_{s}\big]},\notag\\
			\pi_{1,0,1}
			&=\frac{pq\big(1-p_{\alpha^{\text{s}}}p_{s}\big)}{(p+q)\big[p+q+(1-p-q)p_{\alpha^{\text{s}}}p_{s}\big]},\notag\\
			\pi_{0,0,1}&=\pi_{0,1,0}=\pi_{1,0,0}=\pi_{1,1,1}=0.
		\end{align}
		Furthermore, for the MRS policy, we can write
		\begin{align}
			\label{VAoII_pi_ijk_MRS}
			\pi_{0,0,0}
			&=\frac{q\big[q_{\alpha^{\text{s}}_{2}}+p\big(q_{\alpha^{\text{s}}_{1}}-q_{\alpha^{\text{s}}_{2}}\big)\big]\big[q+(1-q)q_{\alpha^{\text{s}}_{2}}p_{s}\big]}{(p+q)F\big(q_{\alpha^{\text{s}}_{1}},q_{\alpha^{\text{s}}_{2}}\big)},\notag\\
			\pi_{0,1,1}
			&=\frac{pq\big(1-q_{\alpha^{\text{s}}_{1}}p_{s}\big)\big[q_{\alpha^{\text{s}}_{2}}+q\big(q_{\alpha^{\text{s}}_{1}}-q_{\alpha^{\text{s}}_{2}}\big)\big]}{(p+q)F\big(q_{\alpha^{\text{s}}_{1}},q_{\alpha^{\text{s}}_{2}}\big)},\notag\\
			\pi_{1,1,0}
			&=\frac{p\big[q_{\alpha^{\text{s}}_{2}}+q\big(q_{\alpha^{\text{s}}_{1}}-q_{\alpha^{\text{s}}_{2}}\big)\big]\big[p+(1-p)q_{\alpha^{\text{s}}_{2}}p_{s}\big]}{(p+q)F\big(q_{\alpha^{\text{s}}_{1}},q_{\alpha^{\text{s}}_{2}}\big)},\notag\\
			\pi_{1,0,1}
			&=\frac{pq\big(1-q_{\alpha^{\text{s}}_{1}}p_{s}\big)\big[q_{\alpha^{\text{s}}_{2}}+p\big(q_{\alpha^{\text{s}}_{1}}-q_{\alpha^{\text{s}}_{2}}\big)\big]}{(p+q)F\big(q_{\alpha^{\text{s}}_{1}},q_{\alpha^{\text{s}}_{2}}\big)},\notag\\
			\pi_{0,0,1}&=\pi_{0,1,0}=\pi_{1,0,0}=\pi_{1,1,1}=0.
		\end{align}
		where $F(\cdot,\cdot)$ in \eqref{VAoII_pi_ijk_MRS} is given by
		\begin{align}
			\label{F_MRS}
			F\big(q_{\alpha^{\text{s}}_{1}},q_{\alpha^{\text{s}}_{2}}\big) &= (1-p)(1-q)p_{s}q^{2}_{\alpha^{\text{s}}_{2}}+(p+q-2pq)q_{\alpha^{\text{s}}_{2}}\notag\\
			&+pq(2-p_{s}q_{\alpha^{\text{s}}_{1}})q_{\alpha^{\text{s}}_{1}}.
		\end{align}
	\end{lemma}
	\begin{IEEEproof}
		See Appendix \ref{Appendix_piijk_VAoII_RS}.
	\end{IEEEproof} 
	
	Now, using \eqref{VAoII_pi_ijk_RS} we can calculate \eqref{PrDAoIVI} for the RS policy as follows
	\begin{align}
		\label{PrDAoIVI2}
		\!\!\!	\mathrm{Pr}\big[{\text{AoIV}}(t)=1\big]&\!\!=\!\pi_{0,1,1}+\pi_{1,0,1}\notag\\
		&\!\!\!=\!\!\frac{2pq\big(1-p_{\alpha^{\text{s}}}p_{s}\big)}{(p\!+\!q)\big[p\!+\!q\!+\!(1\!-\!p\!-\!q)p_{\alpha^{\text{s}}}p_{s}\big]}.
	\end{align}
	Using \eqref{PrDAoIVI2}, the average AoIV, $\overline{\text{AoIV}}$, for the RS policy can be obtained as
	\begin{align}
		\label{Avg_AoIVI_RS}
		\overline{\text{AoIV}} &= \sum_{i=1}^{\infty} i\mathrm{Pr}\big[{\text{AoIV}}(t)=i\big]\notag\\& =  \frac{2pq\big(1-p_{\alpha^{\text{s}}}p_{s}\big)}{(p+q)\big[p+q+(1-p-q)p_{\alpha^{\text{s}}}p_{s}\big]}.
	\end{align}
	Furthermore, using \eqref{VAoII_pi_ijk_MRS}, \eqref{PrDAoIVI} can be written as
	\begin{align}
		\label{PrDAoIVI2_MRS}
		\!\!\!	\mathrm{Pr}\big[{\text{AoIV}}(t)=1\big]&\!\!=\!\pi_{0,1,1}+\pi_{1,0,1}\notag\\
		&\!\!\!=\!\!\frac{pq\big(1\!-\!q_{\alpha^{\text{s}}_{1}}p_{s}\big)\big[(p+q)q_{\alpha^{\text{s}}_{1}}\!+\!(2\!-\!p\!-\!q)q_{\alpha^{\text{s}}_{2}}\big]}{(p+q)F\big(q_{\alpha^{\text{s}}_{1}},q_{\alpha^{\text{s}}_{2}}\big)}.
	\end{align}
	Now, using \eqref{PrDAoIVI2_MRS}, the average AoIV, $\overline{\text{AoIV}}$, for the MRS policy can be calculated as
	\begin{align}
		\label{Avg_AoIVI_MRS}
		\overline{\text{AoIV}} &= \sum_{i=1}^{\infty} i\mathrm{Pr}\big[{\text{AoIV}}(t)=i\big]\notag\\& = \!\!\frac{pq\big(1\!-\!q_{\alpha^{\text{s}}_{1}}p_{s}\big)\big[(p+q)q_{\alpha^{\text{s}}_{1}}\!+\!(2\!-\!p\!-\!q)q_{\alpha^{\text{s}}_{2}}\big]}{(p+q)F\big(q_{\alpha^{\text{s}}_{1}},q_{\alpha^{\text{s}}_{2}}\big)}, 
		\end{align}
	where $F(\cdot,\cdot)$ in \eqref{PrDAoIVI2_MRS} and \eqref{Avg_AoIVI_MRS} is obtained in \eqref{F_MRS}. 
	
	\subsection{Age of Incorrect Information (AoII)}
	\par The AoII quantifies the time elapsed since the transmitted information became incorrect or outdated \cite{maatouk2020age}. Let $\text{AoII}(t)\neq 0$ denote the system being in an erroneous state, i.e., $X(t)\neq \hat{X}(t)$, while the sync state of the system is denoted by $\text{AoII}(t)=0$. We also define $\text{AoII}(t)$ as the AoII at time slot $t$. The evolution of AoII is given by
	\begin{align}
		\label{AoII_definition}
		\text{AoII}\big(t\big) =
		\begin{cases}
			\text{AoII}(t-1)+1, &X(t)\neq \hat{X}(t),\\
			0, &\text{otherwise}.
		\end{cases}
	\end{align}
	\begin{lemma}
		\label{AoIIDistribution_RS}
		For a two-state DTMC information source, $\mathrm{Pr}\big[\text{AoII}(t)=i\big]$ for the RS is given by
		\begin{align}
			&\mathrm{Pr}\big[\text{AoII}(t)=i\big] \notag\\
			&\!\!\!=\!\!\! 
			\begin{cases}
				\frac{p^2+q^2+(p+q-p^2-q^2)p_{\alpha^{\text{s}}}p_{\text{s}}}{(p+q)\big[p+q+(1-p-q)p_{\alpha^{\text{s}}}p_{\text{s}}\big]},&\!\!i\!=\!0,\\
				\frac{pq\big(1-p_{\alpha^{\text{s}}}p_{\text{s}}\big)^{i}\big[(1-q)^{i-1}\Phi(q)+(1-p)^{i-1}\Phi(p)\big]}{(p+q)\big[p+q+(1-p-q)p_{\alpha^{\text{s}}}p_{\text{s}}\big]}, &\!\!i\!\geqslant\! 1.
			\end{cases}
		\end{align}
		where $\Phi(\cdot)$ is given by
		\begin{align}
			\label{Phix}
			\Phi(x) = x+(1-x)p_{\alpha^{\text{s}}}p_{\text{s}}.
		\end{align} 
	\end{lemma}
	\begin{IEEEproof}
		See Appendix \ref{Appendix_AoIIDistribution_RS}.    
	\end{IEEEproof}
	Using Lemma \ref{AoIIDistribution_RS}, we can calculate the average AoII, $\overline{\text{AoII}}$, for the RS policy as follows
	\begin{align}
		\label{Avg_AoII_RS}
		\overline{\text{AoII}} &=\sum_{i=1}^{\infty} i\mathrm{Pr}\big[\text{AoII}(t)\!=\!i\big]\notag\\ &=\frac{pq(1-p_{\alpha^{\text{s}}}p_{s})\big[p+q+(2-p-q)p_{\alpha^{\text{s}}}p_{s}\big]}{(p+q)\Phi(p)\Phi(q)\big[p+q+(1-p-q)p_{\alpha^{\text{s}}}p_{s}\big]}
	\end{align}
	where $\Phi(\cdot)$ is obtained in \eqref{Phix}.  
	\begin{lemma}
		\label{AoIIDistribution_MRS}
		For a two-state DTMC information source, the average AoII, $\overline{\text{AoII}}$, for the MRS policy is given by 
		\begin{align}
			\label{AvgAoII_MRS}
			\!\!\overline{\text{AoII}}\!\!=\!\! \frac{K\big(q_{\alpha^{\text{s}}_{1}},q_{\alpha^{\text{s}}_{2}}\big)}{(p+q)\!\big(q\!+\!(1-q)q_{\alpha^{\text{s}}_{2}}p_{s}\big)\!\big(p\!+\!(1-p)q_{\alpha^{\text{s}}_{2}}p_{s}\big)F\big(q_{\alpha^{\text{s}}_{1}},q_{\alpha^{\text{s}}_{2}}\big)},
		\end{align}
		where $F(\cdot,\cdot)$ given in \eqref{AvgAoII_MRS} is obtained in \eqref{F_MRS} and $K(\cdot,\cdot)$ is given by
		\begin{align}
			\label{K_MRS}
			\!\!\!K\big(q_{\alpha^{\text{s}}_{1}},q_{\alpha^{\text{s}}_{2}}\big) \!\!&=\! pq\big(1-q_{\alpha^{\text{s}}_{1}}p_{s}\big)\bigg[p^{2}\big(q_{\alpha^{\text{s}}_{1}}-q_{\alpha^{\text{s}}_{2}}\big)\big(1-q_{\alpha^{\text{s}}_{2}}p_{s}\big)\notag\\
			\!&+\!pq_{\alpha^{\text{s}}_{2}}\big(1+q_{\alpha^{\text{s}}_{1}}p_{s}-2q_{\alpha^{\text{s}}_{2}}p_{s}\big)+q^2q_{\alpha^{\text{s}}_{1}}\notag\\
			\!&+\!q(1-q)q_{\alpha^{\text{s}}_{2}}\big(1+q_{\alpha^{\text{s}}_{1}}p_{s}\big)\!+\!\big(2-2q+q^{2}\big)p_{s}q^{2}_{\alpha^{\text{s}}_{2}}\bigg].
		\end{align}
	\end{lemma}
	\begin{IEEEproof}
		See Appendix \ref{Appendix_AoIIDistribution_RS}.    
	\end{IEEEproof}
	\par To clarify the difference between VIA, AoIV, and AoII metrics, Fig. \ref{FigCompare} presents an example illustrating the evolution of these metrics. In this figure, we assume that $X(1)=0$ and $\hat{X}(1)=0$. At $t = 3$, the state of the source changes to $X(3)=1$, but there is no successful transmission, i.e., $\hat{X}(3)=0$. Therefore, in this time slot, VIA$(3)=1$, AoIV$(3)=1$, and AoII$(3)=1$. At $t=4$, the state of the source remains unchanged, but the system is in an erroneous state, i.e., $X(4)=1$ and $\hat{X}(4)=0$. Therefore, VIA$(4)=1$, AoIV$(4)=1$, and AoII$(4)=2$. At $t=5$, there is a change in the state of the source $(X(5)=0, \hat{X}(5)=0)$. In this time slot, although the source's state matches the reconstructed state, the transmission fails. Therefore, VIA$(5)=2$, AoIV$(5)=0$, and AoII$(5)=0$. At $t=6$, the source state remains unchanged, and the transmission is successful, i.e., $X(6)=0$, $\hat{X}(6)=0$, resulting in VIA$(6)=0$, AoIV$(6)=0$, and AoII$(6)=0$. Similarly, this explanation can be utilized to analyze the performance of these metrics for $t\geqslant7$.
	\begin{figure}
		\centering 
		
		\begin{tikzpicture}
			\draw [-{Stealth[length=3mm, width=2mm]}](0,0) -- (6,0);
			\draw [-{Stealth[length=3mm, width=2mm]}](0,0) -- (0,3);
			\node(a) at (6,-0.25)  {$t$};
			\node(a1) at (0,-0.25)  {\scriptsize${1}$};
			\node(a1) at (0.5,-0.25)  {\scriptsize${2}$};
			\draw[line width=0.5pt,black] (0.5,-0.05)--(0.5,0.05);
			\node(a1) at (1,-0.25)  {\scriptsize${3}$};
			\draw[line width=0.5pt,black] (1,-0.05)--(1,0.05);
			\node(a1) at (1.5,-0.25)  {\scriptsize${4}$};
			\draw[line width=0.5pt,black] (1.5,-0.05)--(1.5,0.05);
			\node(a1) at (2,-0.25)  {\scriptsize${5}$};
			\draw[line width=0.5pt,black] (2,-0.05)--(2,0.05);
			\node(a1) at (2.5,-0.25)  {\scriptsize${6}$};
			\draw[line width=0.5pt,black] (2.5,-0.05)--(2.5,0.05);
			\node(a1) at (3,-0.25)  {\scriptsize${7}$};
			\draw[line width=0.5pt,black] (3,-0.05)--(3,0.05);
			\node(a1) at (3.5,-0.25)  {\scriptsize${8}$};
			\draw[line width=0.5pt,black] (3.5,-0.05)--(3.5,0.05);
			\node(a1) at (4,-0.25)  {\scriptsize${9}$};
			\draw[line width=0.5pt,black] (4,-0.05)--(4,0.05);
			\node(a1) at (4.5,-0.25)  {\scriptsize${10}$};
			\draw[line width=0.5pt,black] (4.5,-0.05)--(4.5,0.05);
			\node(a1) at (5,-0.25)  {\scriptsize${11}$};
			\node(b1) at (-0.25,0.2)  {\scriptsize${0}$};
			\draw[line width=0.5pt,black] (5,-0.05)--(5,0.05);
			\node(b1) at (-0.25,0.7)  {\scriptsize${1}$};
			\draw[line width=0.5pt,black] (-0.05,0.7)--(0.05,0.7);
			\node(b1) at (-0.25,1.2)  {\scriptsize${2}$};
			\draw[line width=0.5pt,black] (-0.05,1.2)--(0.05,1.2);
			\node(b1) at (-0.25,1.7)  {\scriptsize${3}$};
			\draw[line width=0.5pt,black] (-0.05,1.7)--(0.05,1.7);
			\node(b1) at (-0.25,2)  {\scriptsize${}$};
			\draw[line width=1.7pt, black] (0,0.2) -- (0.5,0.2) -- (1,0.2)--(1,0.7)--(1.5,0.7)--(2,0.7)--(2,1.2)--(2.5,1.2)--(2.5,0.2)--(3,0.2)--(3,0.7)--(3.5,0.
			7)--(3.5,1.2)--(4,1.2)--(4,1.7)--(4.5,1.7)--(5,1.7);
			\draw[line width=0.7pt, red] (0,0.2) -- (1,0.2)--(1,0.7)--(1.5,0.7)--(2,0.7)--(2,0.2)--(3,0.2)--(3,0.7)--(3.5,0.7)--(3.5,0.2)--(4,0.2)--(4,0.7)--(4.5,0.7)--(5,0.7);
			\draw[densely dashed,black,thick] (0,0.2) -- (1,0.2)--(1,0.7)--(1.5,0.7)--(1.5,1.2)--(2,1.2)--(2,0.2)--(3,0.2)--(3,0.7)--(3.5,0.7)--(3.5,0.2)--(4,0.2)--(4,0.7)--(4.5,0.7)--(5,0.7);
			\draw[line width=1.7pt, black] (4,2.6)--(4.5,2.6);
			\node(l1) at (4.8,2.6)  {\scriptsize{VIA}};
			\draw[line width=0.7pt, red] (4,2.35)--(4.5,2.35);
			\node(l2) at (4.87,2.35)  {\scriptsize{AoIV}};
			\draw[densely dashed,black,thick] (4,2.1)--(4.5,2.1);
			\node(l3) at (4.83,2.1)  {\scriptsize{AoII}};
			\draw[line width=0.5pt, black] (3.8,1.9) --(5.25,1.9)--(5.25,2.8)--(3.8,2.8)--(3.8,1.9);
			\node(y1) at (-0.5,2.8)  {\scriptsize{VIA(t)}};
			\node(y2) at (-0.5,2.5)  {\scriptsize{AoIV(t)}};
			\node(y3) at (-0.5,2.2)  {\scriptsize{AoII(t)}};
		\end{tikzpicture}
		\caption{The evolution of the VIA, AoIV, and AoII metrics.}
		\label{FigCompare}
	\end{figure}
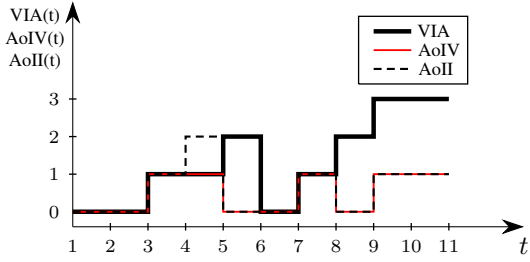
	\begin{remark}
		In what follows, RSC and MRSC policies represent the randomized stationary and modified randomized stationary policies in constrained optimization problems, respectively.	
	\end{remark}
	\section{Constrained Optimization Problems}
	\par In this section, we formulate and solve three constrained optimization problems. In the first one, presented in Section \ref{Optimizationproblem_VIA}, we aim to find the optimal RSC sampling policy that minimizes the average VIA while considering constraints on both the time-averaged sampling cost and the time-averaged reconstruction error, as given in \eqref{PE_RS}. In the second and third optimization problems, presented in Sections \ref{Optimizationproblem_AoIV} and \ref{Optimizationproblem_AoII} respectively, the objective is to find the optimal RSC and MRSC sampling policies that minimize the average AoIV and AoII, respectively, while constraining the time-averaged sampling cost to be less than a certain threshold.
	\vspace{-0.4cm} 
	\subsection{Minimizing the average VIA}
	\label{Optimizationproblem_VIA}
	\par The objective of this optimization problem is to find the optimal value of the probability of sampling in the RS sampling policy, $p_{\alpha^{\text{s}}}$, to minimize the average VIA. We assume that each attempted sampling incurs a cost denoted by $\delta$, and the time-averaged sampling cost is constrained not to surpass a specified threshold $\delta_{\text{max}}$. Furthermore, we consider a constraint that the time-averaged reconstruction error cannot exceed a certain threshold $E_{\text{max}}$. We formulate the following optimization problem 
	\begin{subequations}
		\label{Optimization_problem1}
		\begin{align}
			&\underset{p_{\alpha^{\text{s}}}}{\text{minimize}}\hspace{0.3cm}\overline{\text{VIA}}\label{Optimization_prob1_objfunc}\\
			&\text{subject to}\hspace{0.2cm} \lim_{T \to \infty}\frac{1}{T}\sum_{t=1}^{T}\delta \mathbbm{1}\{\alpha^{\text{s}}(t)=1\} \leqslant\delta_{\text{max}},\label{Optimization_prob1_constraint1}\\
			&\hspace{1.8cm}	P_{\text{E}}\leqslant E_{\text{max}},\label{Optimization_prob1_constraint2}
		\end{align}
	\end{subequations}
	where the constraint in \eqref{Optimization_prob1_constraint1} is the time-averaged sampling cost, which can be simplified as
	\begin{align}
		\label{sampling_cost}
		\lim_{T \to \infty}\frac{1}{T}\sum_{t=1}^{T}\delta \mathbbm{1}\{\alpha^{\text{s}}(t)=1\}=\delta p_{\alpha^{\text{s}}}.
	\end{align}
	Now, using \eqref{Avg_AoIV_RS}, \eqref{PE_RS} and \eqref{sampling_cost} the optimization problem can be formulated as
	\begin{subequations}
		\label{Optimization_problem}
		\begin{align}
			&\!\!\!\!\underset{p_{\alpha^{\text{s}}}}{\text{minimize}}\hspace{0.3cm}\frac{2pq\big(1-p_{\alpha^{\text{s}}}p_{s}\big)}{(p+q)p_{\alpha^{\text{s}}}p_{s}}\label{Optimization_prob_objfunc}\\
			&\!\!\!\!\text{subject to}\hspace{0.2cm}
			\frac{2pq-E_{\text{max}}(p+q)^2}{2pqp_{\text{s}}+E_{\text{max}}(p+q)(1-p-q)p_{\text{s}}}\leqslant p_{\alpha^{\text{s}}}\leqslant \eta,\label{Optimization_prob_constraint}
		\end{align}
	\end{subequations}
	where $\eta = \delta_{\text{max}}/\delta\leqslant 1$ and $ E_{\text{max}}\leqslant 1$.
	\par To solve this optimization problem, we first note that the objective function in \eqref{Optimization_prob_objfunc} is decreasing with $p_{\alpha^{\text{s}}}$. In other words, the objective function has its minimum value when $p_{\alpha^{\text{s}}}$ has its maximum. Using the constraint given in \eqref{Optimization_prob_constraint}, the maximum value of sampling probability is $\eta$. However, $\eta$ is the optimal value of sampling probability when $ \frac{2pq-E_{\text{max}}(p+q)^2}{2pqp_{\text{s}}+E_{\text{max}}(p+q)(1-p-q)p_{\text{s}}}\leqslant\eta\leqslant 1$; otherwise, we cannot find a sampling probability that satisfies the constraint of the optimization problem, and thus, an optimal solution does not exist.
	\vspace{-0.4cm}
	\subsection{Minimizing the average AoIV}
	\label{Optimizationproblem_AoIV}
	\par The goal of this optimization problem is to determine the optimal values for the sampling probabilities, $p_{\alpha^{\text{s}}}$, $q_{\alpha^{\text{s}}_{1}}$, and $q_{\alpha^{\text{s}}_{2}}$ for both the RSC and MRSC sampling policies, respectively, which minimize the average AoIV while considering the time-averaged sampling cost constraint. Here, $\delta$ represents the cost of each attempted sampling, and $\delta_{\text{max}}$ is the total average cost. Using \eqref{Avg_AoIVI_RS}, this optimization problem for the RSC policy is formulated as follows:
	\begin{subequations}
		\label{Optimization_problemAoIV}
		\begin{align}
			&\underset{p_{\alpha^{\text{s}}}}{\text{minimize}}\hspace{0.3cm}\frac{2pq\big(1-p_{\alpha^{\text{s}}}p_{s}\big)}{(p+q)\big[p+q+(1-p-q)p_{\alpha^{\text{s}}}p_{s}\big]}\label{Optimization_probAoIV_objfunc}\\
			&\text{subject to}\hspace{0.2cm} \lim_{T \to \infty}\frac{1}{T}\sum_{t=1}^{T}\delta \mathbbm{1}\{\alpha^{\text{s}}(t)=1\} \leqslant\delta_{\text{max}}.\label{Optimization_probAoIV_constraint}
		\end{align}
	\end{subequations}
	\par To solve this optimization problem, we note that the objective function given in \eqref{Optimization_probAoIV_objfunc} decreases as $p_{\alpha^{\text{s}}}$ increases. In other words, to minimize the objective function, we must find the maximum value of $p_{\alpha^{\text{s}}}$ that satisfies the constraint in \eqref{Optimization_probAoIV_constraint}. Using \eqref{sampling_cost}, for the RSC policy, the maximum value of the sampling probability that minimizes the average AoIV is given by $p^{*}_{\alpha^{\text{s}}}=\eta$, where $0 \leqslant \eta = \delta_{\text{max}}/\delta \leqslant 1$. Now, using \eqref{Avg_AoIVI_MRS} for the MRSC policy, we formulate the optimization problem as follows
	\begin{subequations}		
		\label{Optimization_problem2_AoIV_MRS}
		\begin{align}	&\!\!\underset{q_{\alpha^{\text{s}}_{1}},q_{\alpha^{\text{s}}_{2}}}{\text{minimize}}\hspace{0.2cm}\frac{pq\big(1\!-\!q_{\alpha^{\text{s}}_{1}}p_{s}\big)\big[(p+q)q_{\alpha^{\text{s}}_{1}}\!+\!(2\!-\!p\!-\!q)q_{\alpha^{\text{s}}_{2}}\big]}{(p+q)F\big(q_{\alpha^{\text{s}}_{1}},q_{\alpha^{\text{s}}_{2}}\big)}\label{Optimization_prob2_AoIV_MRS_objfunc}\\
			&\text{subject to}\hspace{0.2cm} \lim_{T \to \infty}\frac{1}{T}\sum_{t=1}^{T}\delta \mathbbm{1}\{\alpha^{\text{s}}(t)=1\} \leqslant\delta_{\text{max}},\label{Optimization_prob2_AoIV_MRS_constraint}
		\end{align}
	\end{subequations}
	where $F(\cdot,\cdot)$ is given in \eqref{F_MRS}. Since this optimization problem is not convex with respect to $q_{\alpha^{\text{s}}_{1}}$ and $q_{\alpha^{\text{s}}_{2}}$, and the optimization parameters are interdependent, a closed-form solution is hard or impossible to be found for global optimization. Therefore, we solve \eqref{Optimization_problem2_AoIV_MRS} numerically. Furthermore, as a special case, by using the following Lemma, we can obtain the optimal value of the probability of sampling for the MRSC policy when $q_{\alpha^{\text{s}}_{1}}=q_{\alpha^{\text{s}}_{2}}$.
	\begin{lemma}
		\label{OptimalMRS}
		In the MRSC sampling policy, when $q_{\alpha^{\text{s}}_{1}}=q_{\alpha^{\text{s}}_{2}} = q_{\alpha^{\text{s}}}$ the optimal value of the sampling probability, i.e., $q^{*}_{\alpha^{\text{s}}}$ is given by
		\begin{align}
			&q^{*}_{\alpha^{\text{s}}}\notag\\
			&\!=\! 
			\begin{cases}
				1,&\!\!\!\!\eta(p\!+\!q)\!(1\!-\!p\!-\!q)p_{s}\!\!\geqslant\!\! 2pq,\notag\\
				\!\text{min}\left\{\!1,\frac{\eta(p+q)^{2}}{2pq-\eta(p+q)\!(1-p-q)p_{s}}\!\right\}\!,&\!\!\!\!\eta(p\!+\!q)\!(1\!-\!p\!-\!q)p_{s}\!\!<\!\! 2pq.
			\end{cases}
		\end{align}
		\begin{IEEEproof}
			See Appendix \ref{Appendix_MRS_ASamplingCost}.    
		\end{IEEEproof}
	\end{lemma}
	
	\subsection{Minimizing the average AoII}
	\label{Optimizationproblem_AoII}
	\par The objective of this optimization problem is to obtain the optimal values of the sampling probabilities, $p_{\alpha^{\text{s}}}$, $q_{\alpha^{\text{s}}_{1}}$, and $q_{\alpha^{\text{s}}_{2}}$, for both the RSC and MRSC sampling policies, respectively. The aim is to minimize the average AoII while ensuring that the time-averaged sampling cost remains below a certain threshold. Let $\delta$ represent the cost of each attempted sampling and $\delta_{\text{max}}$ denote the total average cost. Using the expression given in \eqref{Avg_AoII_RS}, we formulate this optimization problem for the RSC policy as follows:
	\begin{subequations}
		\label{Optimization_problem2}
		\begin{align}
			&\underset{p_{\alpha^{\text{s}}}}{\text{minimize}}\hspace{0.3cm}{\frac{pq(1-p_{\alpha^{\text{s}}}p_{s})\big[p+q+(2-p-q)p_{\alpha^{\text{s}}}p_{s}\big]}{(p+q)\Phi(p)\Phi(q)\big[p+q+(1-p-q)p_{\alpha^{\text{s}}}p_{s}\big]}}\label{Optimization_prob2_objfunc}\\
			&\text{subject to}\hspace{0.2cm} \lim_{T \to \infty}\frac{1}{T}\sum_{t=1}^{T}\delta \mathbbm{1}\{\alpha^{\text{s}}(t)=1\} \leqslant\delta_{\text{max}},\label{Optimization_prob2_constraint}
		\end{align}
	\end{subequations}
	where $\Phi(\cdot)$ in \eqref{Optimization_prob2_objfunc} is obtained in \eqref{Phix}. We note that the objective function in \eqref{Optimization_prob2_objfunc} decreases with $p_{\alpha^{\text{s}}}$. Therefore, to solve this optimization problem, we need to determine the maximum value of $p_{\alpha^{\text{s}}}$ that satisfies the constraint given in \eqref{Optimization_prob2_constraint}. Using \eqref{sampling_cost}, for the RSC policy, the maximum value of the probability of sampling that minimizes the average AoII is given by $p^{*}_{\alpha^{\text{s}}}=\eta$, where $0\leqslant\eta=\delta_{\text{max}}/\delta\leqslant 1$. Using \eqref{AvgAoII_MRS}, we formulate the optimization problem for the MRSC policy as follows:
	
	\begin{subequations}
		\label{Optimization_problem3_AoII_MRS}
		\begin{align}	&\!\!\underset{q_{\alpha^{\text{s}}_{1}},q_{\alpha^{\text{s}}_{2}}}{\text{minimize}}\frac{K\big(q_{\alpha^{\text{s}}_{1}},q_{\alpha^{\text{s}}_{2}}\big)}{(p\!+\!q)\!\big(q\!+\!(1\!-\!q)q_{\alpha^{\text{s}}_{2}}p_{s}\big)\!\big(p\!+\!(1\!-\!p)q_{\alpha^{\text{s}}_{2}}p_{s}\big)\!F\big(q_{\alpha^{\text{s}}_{1}}\!,\!q_{\alpha^{\text{s}}_{2}}\big)}\label{Optimization_prob3_MRS_objfunc}\\
			&\text{subject to}\hspace{0.2cm} \lim_{T \to \infty}\frac{1}{T}\sum_{t=1}^{T}\delta \mathbbm{1}\{\alpha^{\text{s}}(t)=1\} \leqslant\delta_{\text{max}},\label{Optimization_prob3_MRS_constraint}
		\end{align}
	\end{subequations}
	where $F(\cdot,\cdot)$ and $K(\cdot,\cdot)$ are given in \eqref{F_MRS} and \eqref{K_MRS}, respectively. This optimization problem is not convex with respect to $q_{\alpha^{\text{s}}_{1}}$ and $q_{\alpha^{\text{s}}_{2}}$, and the optimization parameters are interdependent. Therefore, obtaining a closed-form solution for global optimization is cumbersome or infeasible, and we solve \eqref{Optimization_problem3_AoII_MRS} numerically. Furthermore, as a special case, when $q_{\alpha^{\text{s}}_{1}}=q_{\alpha^{\text{s}}_{2}} = q_{\alpha^{\text{s}}}$, the optimal value of the probability of sampling, denoted as $q^{*}_{\alpha^{\text{s}}}$, that minimizes the average AoII is given by Lemma \ref{OptimalMRS}.
	
	\vspace{-0.3cm}
	\section{Numerical Results}
	\par In this section, we numerically validate our analytical results and assess the performance of the sampling policies in terms of average VIA, average AoIV, and average AoII under various system parameters. Simulation results are obtained by averaging over $10^7$ time slots, and the initial values for the state of the source and the reconstructed source are set to $X(1) = 0$ and $\hat{X}(1) = 0$, respectively.
	\begin{figure}[!t]
		\centering
		\subfigure[$p_{{\text{s}}} = 0.3$ ]{\includegraphics[trim=0.5cm 0.05cm 1.1cm 0.6cm,width=0.48\linewidth, clip]{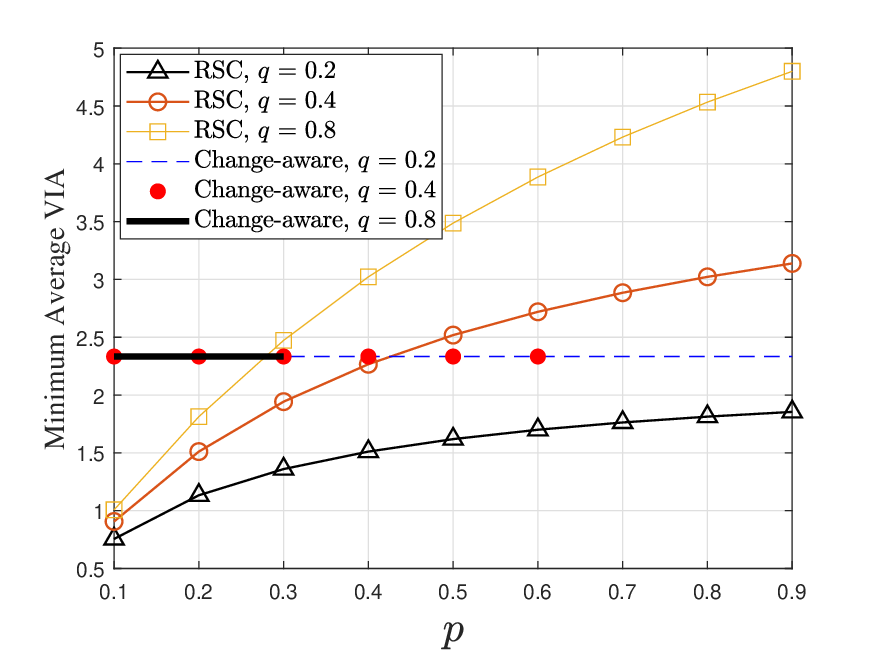}
		}
		\subfigure[$p_{{\text{s}}} = 0.7$]{\centering
			\includegraphics[trim=0.5cm 0.05cm 1.1cm 0.6cm,width=0.48\linewidth, clip]{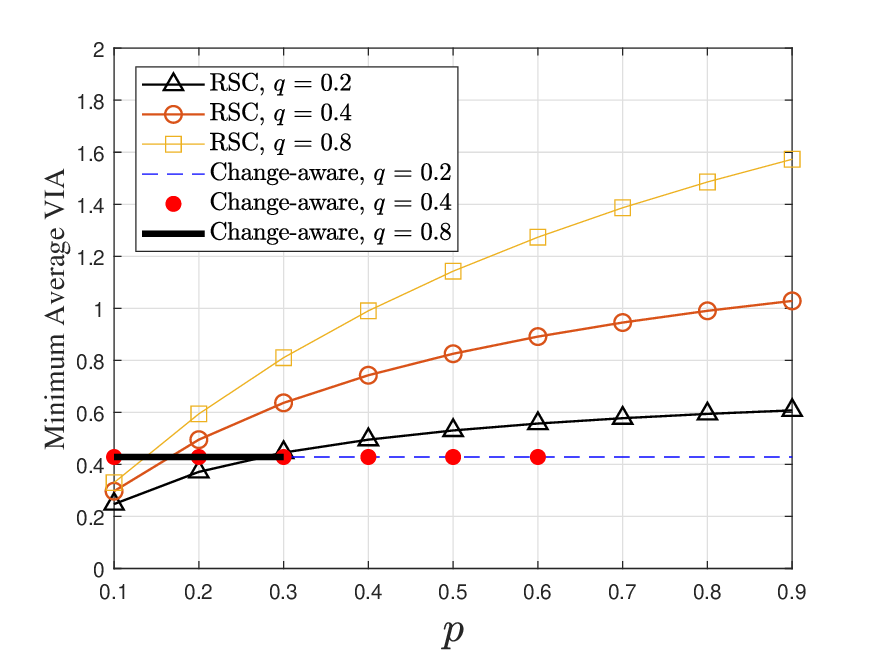}
		}
		\caption{Minimum Average VIA in a constrained optimization problem as a function of $p$ and $q$.}
		\label{Min_VAoI_Constrained}
	\end{figure}
	\begin{figure}[!t]
		\centering
		\subfigure[$p_{{\text{s}}} = 0.3$ ]{\includegraphics[trim=0.5cm 0.05cm 1.1cm 0.6cm,width=0.48\linewidth, clip]{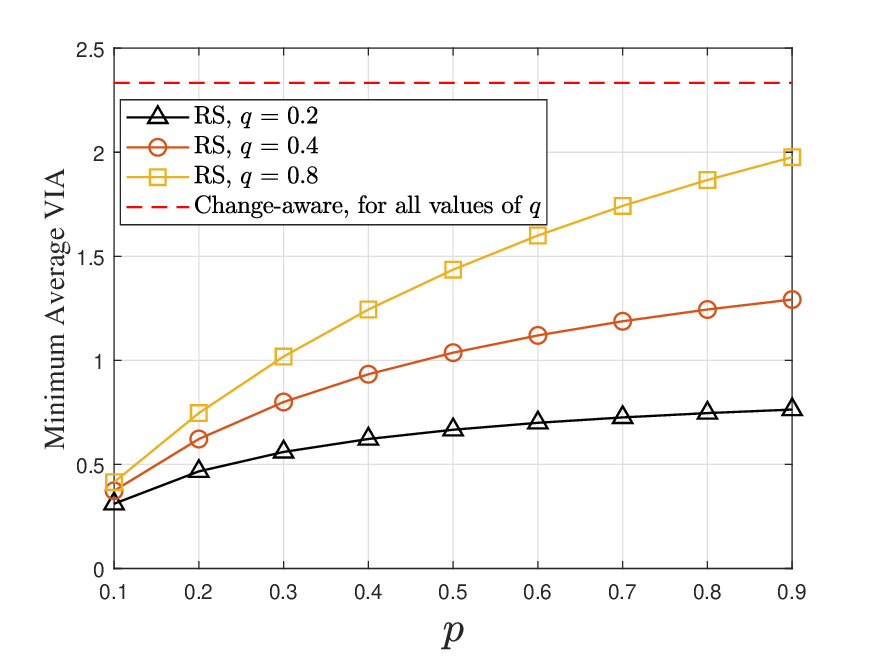}
		}
		\subfigure[$p_{{\text{s}}} = 0.7$]{\centering
			\includegraphics[trim=0.5cm 0.05cm 1.1cm 0.6cm,width=0.48\linewidth, clip]{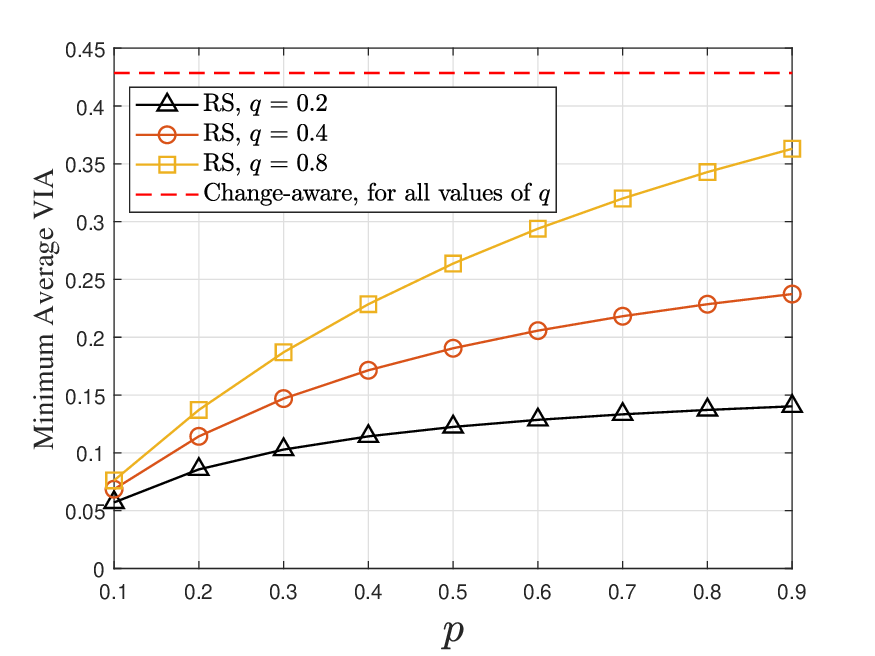}
		}
		\caption{Minimum Average VIA in an unconstrained optimization problem as a function of $p$ and $q$.}
		\label{Min_VAoI_UnConstrained}
	\end{figure}
	
	Figs. \ref{Min_VAoI_Constrained} and \ref{Min_VAoI_UnConstrained} illustrate the minimum average VIA under time-averaged sampling cost and reconstruction error constraints for $\eta = 0.5$ and $E_{\text{max}} = 0.5$, considering various values of $p$, $q$, and $p_{\text{s}}$ for both the constrained and unconstrained optimization problems, respectively. As seen in Fig. \ref{Min_VAoI_Constrained}, with both low and high success probabilities, the optimal RSC outperforms the change-aware policy in scenarios where the source changes slowly and rapidly. In contrast, the change-aware policy exhibits superior performance for a moderately changing source. This is because when the source changes slowly, the change-aware policy takes fewer sampling and transmission actions, resulting in a higher average VIA. In contrast, when the source changes rapidly, the change-aware policy generates more samples, violating the imposed constraint, and in that case, the optimal RSC performs better. Moreover, according to Remark \ref{remark_RS_CA_compare_AoIV}, the RS policy achieves a lower average VIA compared to the change-aware policy when $\frac{2pq}{p+q+\big(2pq-p-q\big)p_{\text{s}}}\leqslant p_{\alpha^{\text{s}}}<1$. However, based on the constraint given in \eqref{Optimization_prob_constraint}, the maximum sampling probability is limited to $\eta$. Consequently, for values of $p$ and $q$ where $\eta<\frac{2pq}{p+q+\big(2pq-p-q\big)p_{\text{s}}}$, the change-aware policy outperforms the optimal RSC policy in terms of average VIA. Furthermore, as shown in Fig. \ref{Min_VAoI_UnConstrained}, the optimal performance of the RS policy surpasses that of the change-aware policy in the unconstrained case. However, in such a scenario, the optimal strategy for the RS policy involves sampling at every time slot. 
	Moreover, transmitting at every time slot results in the generation of an excessive amount of samples.
	\begin{figure}[!t]
		\centering
		
		\subfigure[$p_{{\text{s}}} = 0.1$ ]{\includegraphics[trim=0.5cm 0.05cm 1.1cm 0.6cm,width=0.48\linewidth, clip]{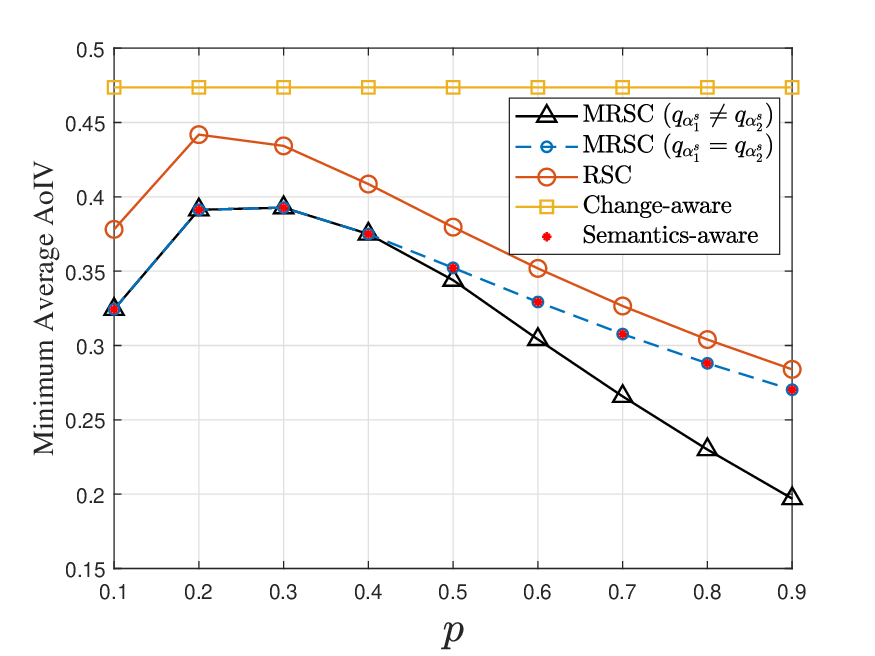}
			\label{MinAvg_AoIV_q0.2_ps0.1}
		}
		\subfigure[$p_{{\text{s}}} = 0.9$]{\centering
			\includegraphics[trim=0.5cm 0.05cm 1.1cm 0.6cm,width=0.48\linewidth, clip]{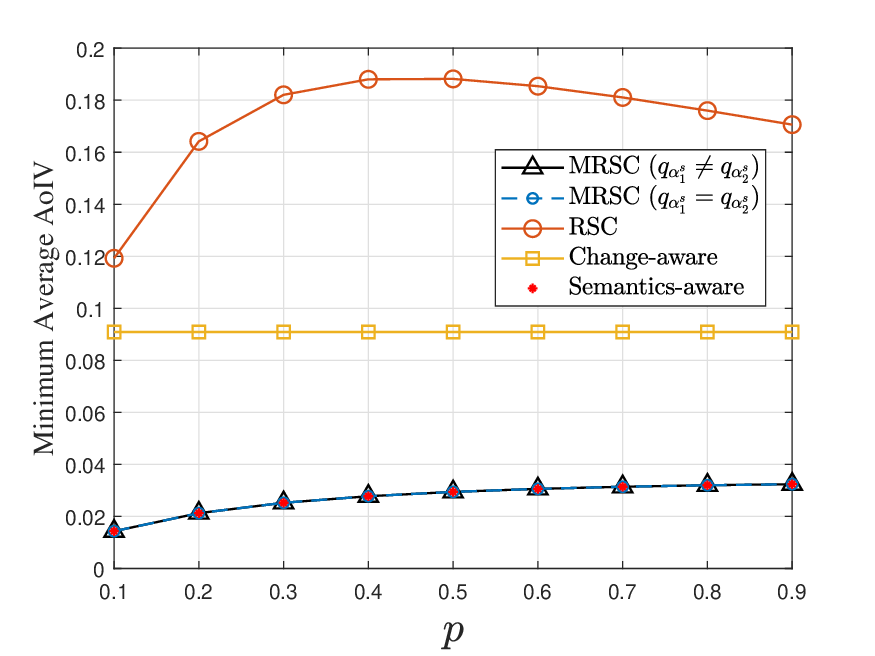}
		}
		\caption{Minimum Average AoIV as a function of $p$ for $q = 0.2$ and $\eta = 0.5$.}
		\label{MinAvg_AoIV_q0.2}
	\end{figure}
	\begin{figure}[!t]
		\centering
		\subfigure[$p_{{\text{s}}} = 0.1$ ]{\includegraphics[trim=0.5cm 0.05cm 1.1cm 0.6cm,width=0.48\linewidth, clip]{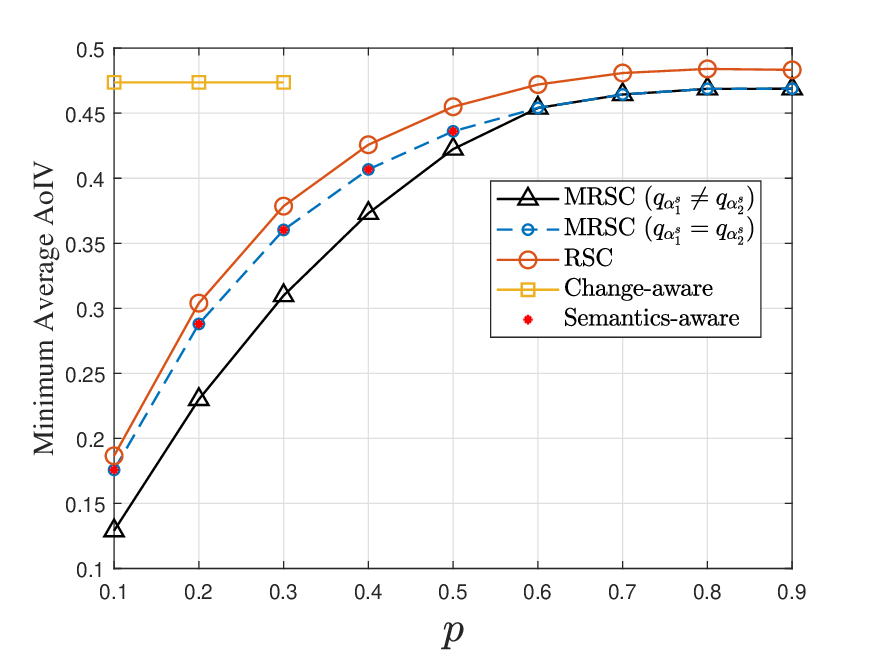}
			\label{MinAvg_AoIV_q0.8_ps0.1}
		}
		\subfigure[$p_{{\text{s}}} = 0.9$ ]{\includegraphics[trim=0.5cm 0.05cm 1.1cm 0.6cm,width=0.48\linewidth, clip]{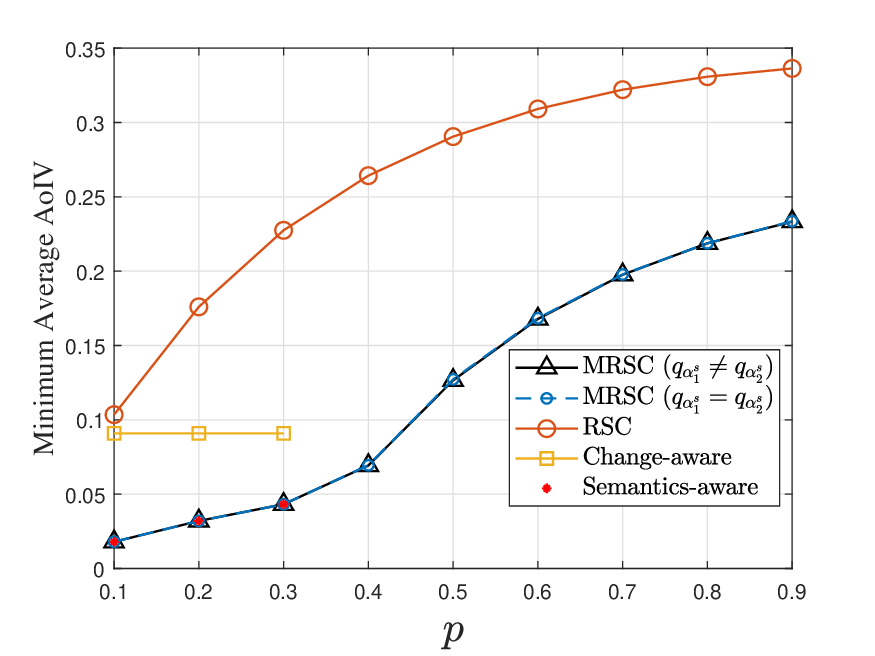}
		}
		\caption{Minimum Average AoIV as a function of $p$ for $q = 0.8$ and $\eta = 0.5$.}
		\label{MinAvg_AoIV_q0.8}
	\end{figure}
	\par
\begin{figure}[!t]
		\centering
		
		\subfigure[$p_{{\text{s}}} = 0.1$ ]{\includegraphics[trim=0.5cm 0.05cm 1.1cm 0.6cm,width=0.48\linewidth, clip]{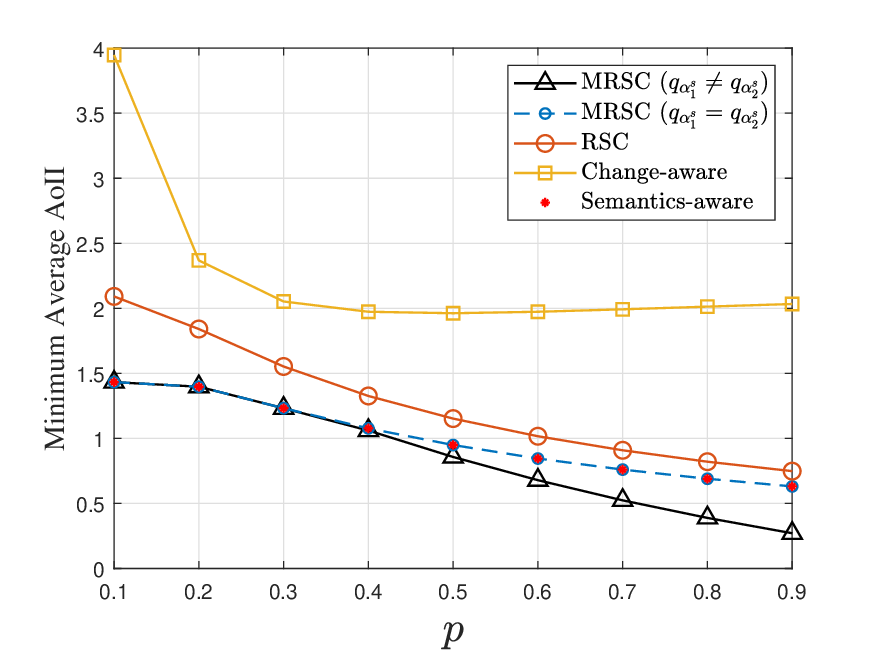}
			\label{MinAvgAoII_ps0.1q0.2}}
		\subfigure[$p_{{\text{s}}} = 0.9$]{\centering
			\includegraphics[trim=0.5cm 0.05cm 1.1cm 0.6cm,width=0.48\linewidth, clip]{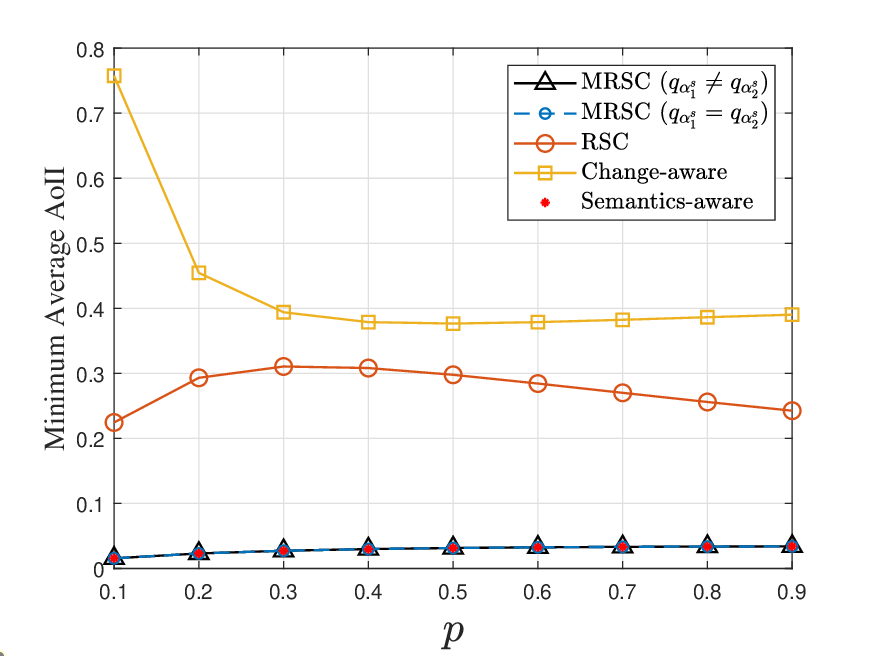}
		}
		\caption{Minimum Average AoII as a function of $p$ for $q = 0.2$ and $\eta = 0.5$.}
		\label{MinAvgAoII_q0.2}
	\end{figure}

\begin{figure}[!t]
		\centering
		\subfigure[$p_{{\text{s}}} = 0.1$ ]{\includegraphics[trim=0.5cm 0.05cm 1.1cm 0.6cm,width=0.48\linewidth, clip]{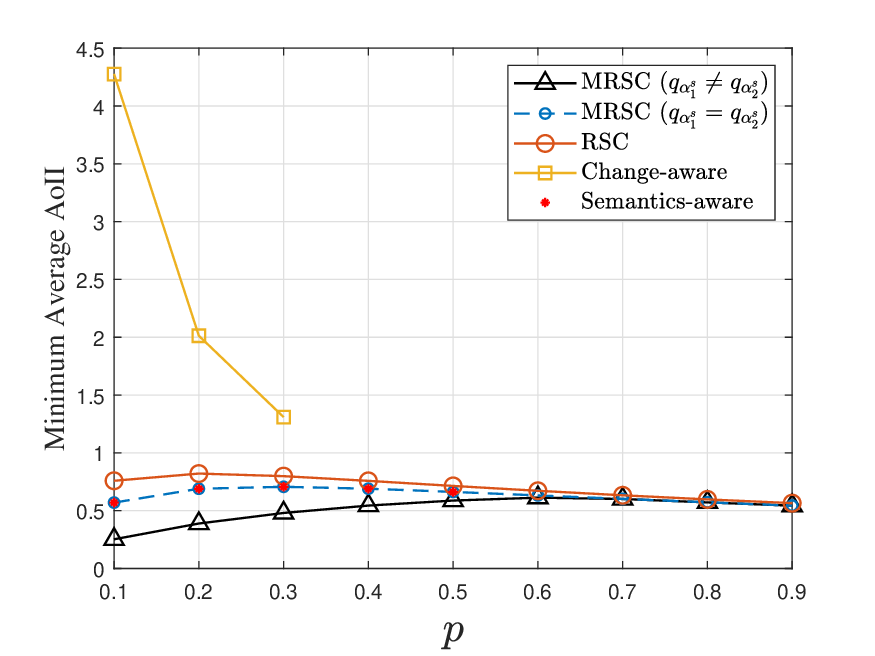}
			\label{MinAvgAoII_ps0.1q0.8}}
		\subfigure[$p_{{\text{s}}} = 0.9$]{\centering
			\includegraphics[trim=0.5cm 0.05cm 1.1cm 0.6cm,width=0.48\linewidth, clip]{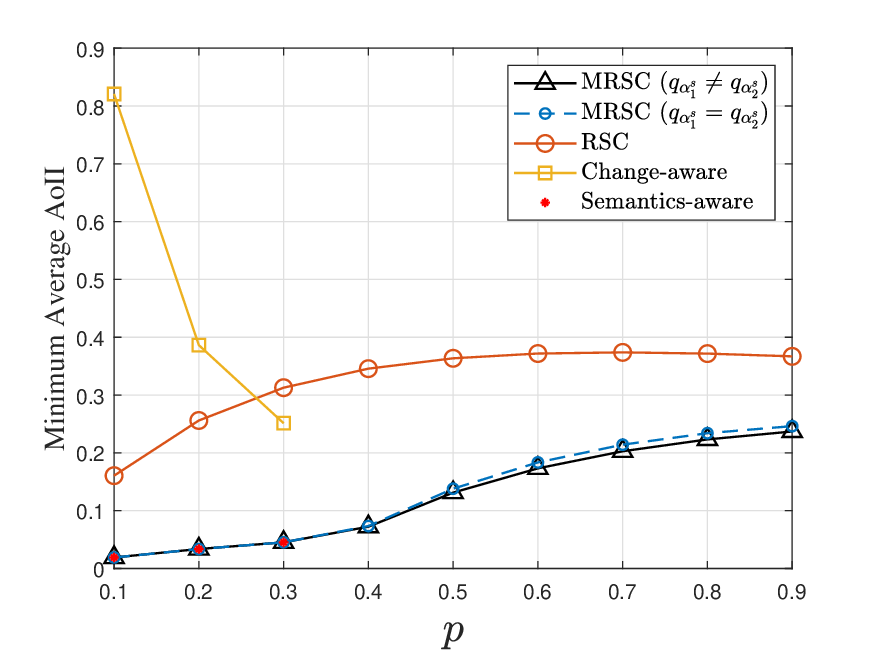}
		}
		\caption{Minimum Average AoII as a function of $p$ for $q = 0.8$ and $\eta = 0.5$.}
		\label{MinAvgAoII_q0.8}
	\end{figure}

\begin{figure}[!t]
		\centering
		\subfigure[$p_{{\text{s}}} = 0.1$ ]{\includegraphics[trim=0.5cm 0.05cm 1.1cm 0.7cm,width=0.48\linewidth, clip]{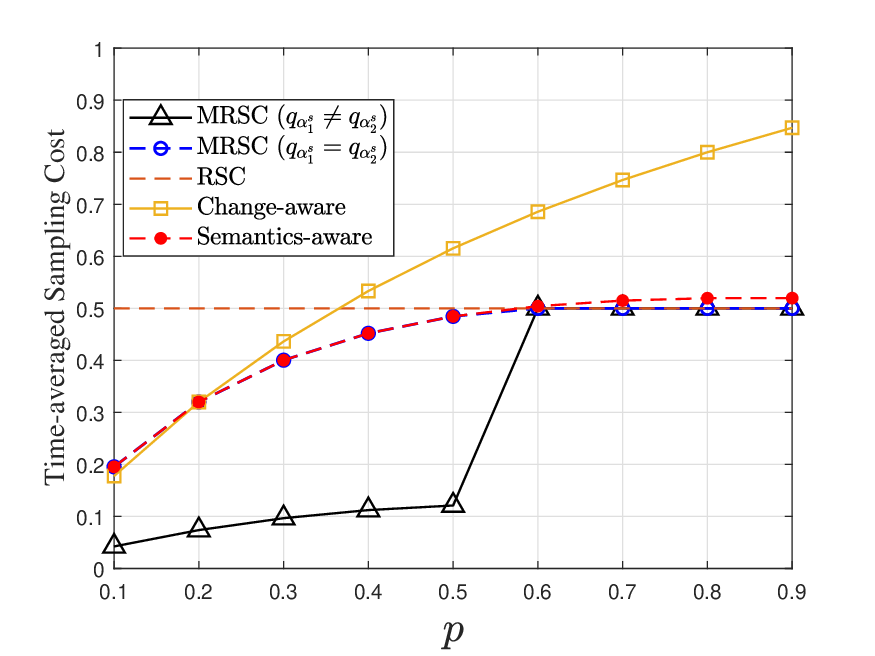}
			\label{SamplinCostAvgAoII_ps0.1q0.8}}
		\subfigure[$p_{{\text{s}}} = 0.9$]{\centering
			\includegraphics[trim=0.5cm 0.05cm 1.1cm 0.7cm,width=0.48\linewidth, clip]{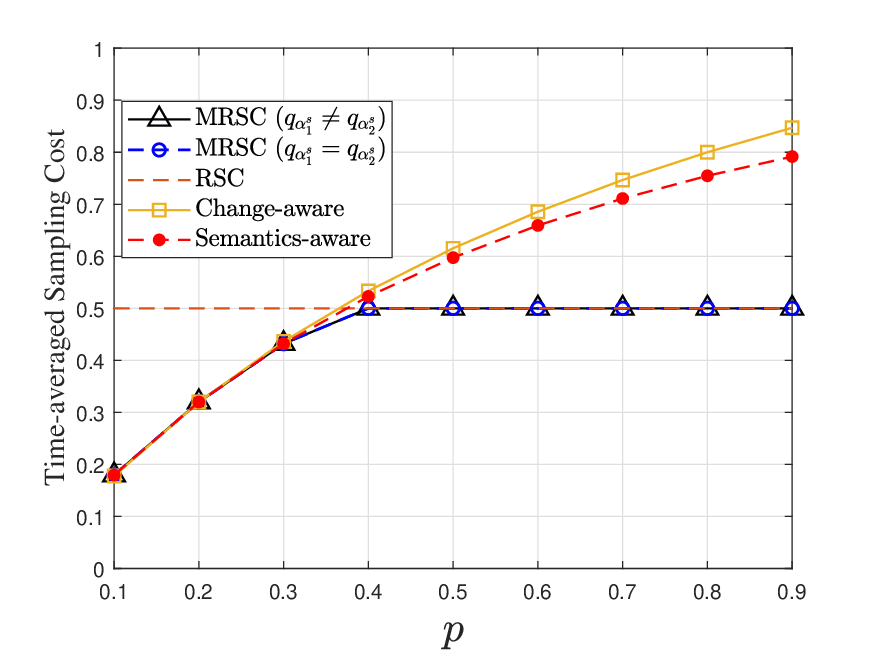}
		}
		\caption{Time-averaged sampling cost as a function of $p$ for $q = 0.8$ and $\eta = 0.5$.}
		\label{SamplinCostAvgAoII_q0.8}
	\end{figure}

 Figs. \ref{MinAvg_AoIV_q0.2} and \ref{MinAvg_AoIV_q0.8} illustrate the minimum average AoIV  and Figs. \ref{MinAvgAoII_q0.2} and \ref{MinAvgAoII_q0.8} depict the minimum average AoII under a time-averaged sampling cost constraint as a function of $p$ for $\eta = 0.5$, and selected values of $p_{s}$ and $q$. We observe that the optimal MRSC, regardless of whether $q_{\alpha^{\text{s}}_{1}}$ is equal to $q_{\alpha^{\text{s}}_{2}}$ or not, outperforms all other policies across scenarios of slow, moderate, and rapid source state changes. The reason behind this is that in the MRSC policy, at time slot $t$, we perform sampling \emph{probabilistically}, only when $X(t)\neq \hat{X}(t-1)$. Therefore, this policy requires fewer sampling actions than the change-aware and semantics-aware policies, which in turn can violate the imposed constraint for rapidly changing sources. Furthermore, in the optimal MRSC policy, we can perform sampling with a higher probability compared to the optimal RSC policy while incurring either fewer or, in the worst case, the same time-averaged sampling cost for a rapidly changing source, as shown in Fig. \ref{SamplinCostAvgAoII_q0.8}. Therefore, the optimal MRSC has a lower average AoIV and AoII than the optimal RSC policy. Moreover, using Tables \ref{Table3_MRSC} and \ref{Table6_MRSC}, for the MRSC, varying the sampling probabilities results in lower average AoIV and AoII compared to when the same sampling probabilities are considered, particularly for a lower success probability and when the source remains in one state with high probability. According to these tables, when the total value of time-averaged sampling cost is limited, considering a higher value for $q_{\alpha^{\text{s}}_{2}}$ and a lower value for $q_{\alpha^{\text{s}}_{1}}$ results in lower average AoIV and AoII compared to the case when we utilize the same but lower sampling probabilities. Refraining from sampling when the system is in an erroneous state is more detrimental than when the system is in sync and no sampling is performed. Therefore, it is reasonable to consider $q_{\alpha^{\text{s}}_{2}}\geqslant q_{\alpha^{\text{s}}_{1}}$. Furthermore, when the source remains in one state with a high probability and the success probability is low, the optimal policy for MRSC is never to perform sampling when the system is in sync and always perform sampling when the system is in an erroneous state. In this scenario, MRSC generates fewer samples and results in lower average AoIV and AoII compared to other policies.
	
	\begin{table}[!h]
		\centering
		\caption{{Optimal values of the sampling probabilities for RSC and MRSC that minimize average AoIV and AoII for $p_{\text{s}}= 0.1$, $\eta = 0.5$. 
		}}
		\subtable[$q = 0.2$]{
			\label{Table1_MRSC}
			\begin{tabular}{|c|c|c|c|c|c|}
				\hline
				$p$ & 0.1 &0.3&0.5&0.7&0.9\\ \hline
				$p^{*}_{\alpha^{\text{s}}}$&0.5&0.5&0.5&0.5& 0.5\\\hline
				$q^{*}_{\alpha^{\text{s}}_{1}}$&1&1&0&0&0\\\hline
				$q^{*}_{\alpha^{\text{s}}_{2}}$&1&1&1&1&1\\\hline
				$q^{*}_{\alpha^{\text{s}}}$&1&1&1&1&1\\\hline
			\end{tabular}
		}
		\hfill
		\subtable[$q = 0.8$]{
			\label{Table2_MRSC}
			\begin{tabular}{|c|c|c|c|c|c|}
				\hline
				$p$ & 0.1 &0.3&0.5&0.7&0.9\\ \hline
				$p^{*}_{\alpha^{\text{s}}}$&0.5&0.5&0.5&0.5& 0.5\\\hline
				$q^{*}_{\alpha^{\text{s}}_{1}}$&0&0&0&0.963&0.958\\\hline
				$q^{*}_{\alpha^{\text{s}}_{2}}$&1&1&1&1&1\\\hline
				$q^{*}_{\alpha^{\text{s}}}$&1&1&1&0.972&0.963\\\hline
			\end{tabular}
		}
		\label{Table3_MRSC}
	\end{table}
	\begin{table}[!h]
		\centering
		\caption{{Optimal values of the sampling probabilities for both RSC and MRSC that minimize average AoIV and AoII for $p_{\text{s}}= 0.9$, $\eta = 0.5$.
		}}
		\subtable[$q = 0.2$]{
			\label{Table4_MRSC}
			\begin{tabular}{|c|c|c|c|c|c|}
				\hline
				$p$ & 0.1 &0.3&0.5&0.7&0.9\\ \hline
				$p^{*}_{\alpha^{\text{s}}}$&0.5&0.5&0.5&0.5& 0.5\\\hline
				$q^{*}_{\alpha^{\text{s}}_{1}}$&1&1&1&1&1\\\hline
				$q^{*}_{\alpha^{\text{s}}_{2}}$&1&1&1&1&1\\\hline
				$q^{*}_{\alpha^{\text{s}}}$&1&1&1&1&1\\\hline
			\end{tabular}
		}
		\hfill
		\subtable[$q = 0.8$]{
			\label{Table5_MRSC}
			\begin{tabular}{|c|c|c|c|c|c|}
				\hline
				$p$ & 0.1 &0.3&0.5&0.7&0.9\\ \hline
				$p^{*}_{\alpha^{\text{s}}}$&0.5&0.5&0.5&0.5& 0.5\\\hline
				$q^{*}_{\alpha^{\text{s}}_{1}}$&1&1&0.856&0.753&0.717\\\hline
				$q^{*}_{\alpha^{\text{s}}_{2}}$&1&1&1&1&1\\\hline
				$q^{*}_{\alpha^{\text{s}}}$&1&1&0.866&0.772&0.731\\\hline
			\end{tabular}
		}
		\label{Table6_MRSC}
	\end{table}
	\begin{figure}[!t]
		\centering
		\subfigure[$p_{{\text{s}}} = 0.1$ ]{\includegraphics[trim=0.5cm 0.05cm 1.1cm 0.7cm,width=0.48\linewidth, clip]{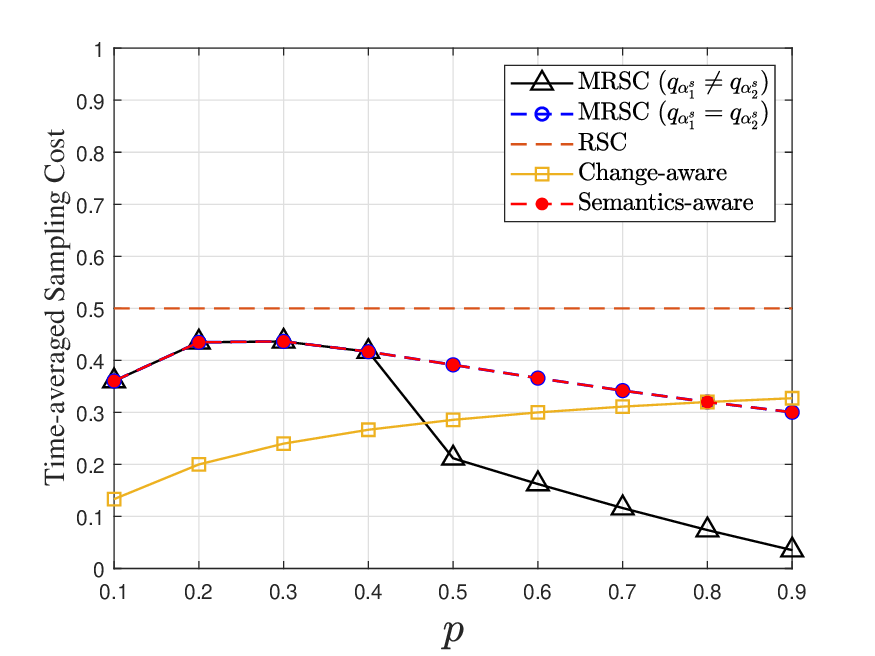}
			\label{SamplinCostAvgAoII_ps0.1q0.2}}
		\subfigure[$p_{{\text{s}}} = 0.9$]{\centering
			\includegraphics[trim=0.5cm 0.05cm 1.1cm 0.7cm,width=0.48\linewidth, clip]{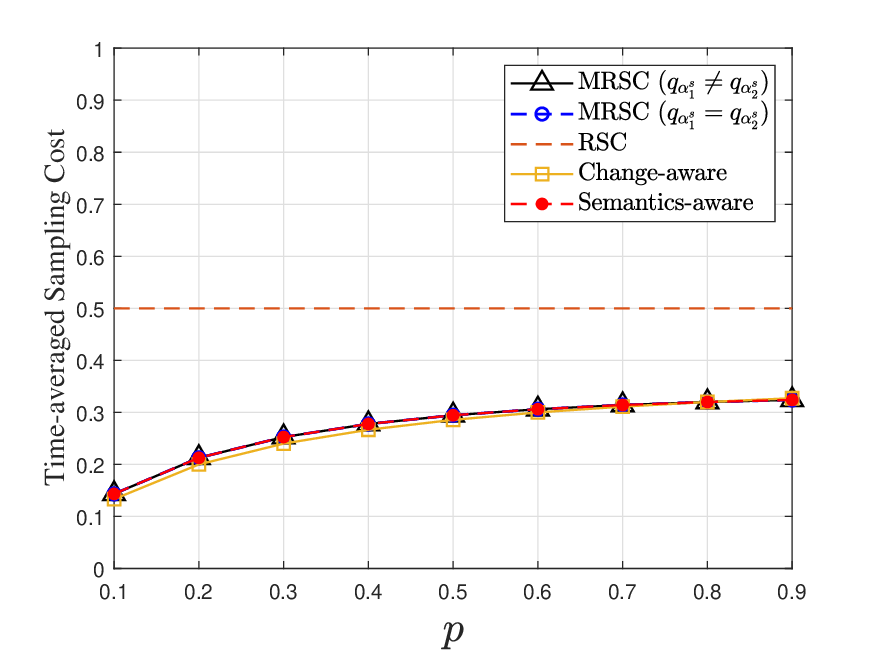}
		}
		\caption{Time-averaged sampling cost as a function of $p$ for $q = 0.2$ and $\eta = 0.5$.}
		\label{SamplinCostAvgAoII_q0.2}
	\end{figure}
	\par Figs. \ref{SamplinCostAvgAoII_q0.2} and \ref{SamplinCostAvgAoII_q0.8} illustrate the time-averaged sampling cost as a function of $p$ for $\eta = 0.5$ and selected values of $q$ and $p_{s}$. In these figures, the time-averaged sampling cost for the MRSC is obtained for values of $q_{\alpha^{\text{s}}_{1}}$ and $q_{\alpha^{\text{s}}_{2}}$ that minimize the average AoIV and AoII. These figures show that when both $p$ and $q$ have high values (indicating rapid source changes), the semantics-aware and change-aware policies exhibit higher time-averaged sampling costs than other policies. Therefore, these values of $p$ and $q$ can violate the time-averaged sampling cost constraint, and the constrained optimization problems do not have optimal solutions, as shown in Figs. \ref{MinAvg_AoIV_q0.8} and \ref{MinAvgAoII_q0.8}. However, under these values of $p$ and $q$, the optimal MRSC policy ensures that the maximum value of the time-averaged sampling cost equals the specified threshold ($\eta$) of the time-averaged sampling cost constraint. Furthermore, using Figs. \ref {MinAvg_AoIV_q0.2_ps0.1}, \ref{MinAvg_AoIV_q0.8_ps0.1}, \ref{MinAvgAoII_ps0.1q0.2}, \ref{MinAvgAoII_ps0.1q0.8}, \ref{SamplinCostAvgAoII_ps0.1q0.2}, and \ref{SamplinCostAvgAoII_ps0.1q0.8}, for a lower success probability ($p_{s} = 0.1$) and when only one of $p$ or $q$ has a high value (indicating that the source remains in state $0$ or $1$ with high probability), the MRSC not only generates fewer samples but also results in a lower average AoIV and AoII.
	\begin{figure}[!h]
		\centering
		\subfigure[$p_{{\text{s}}} = 0.1$ ]{\includegraphics[trim=0.5cm 0.05cm 1.1cm 0.7cm,width=0.48\linewidth, clip]{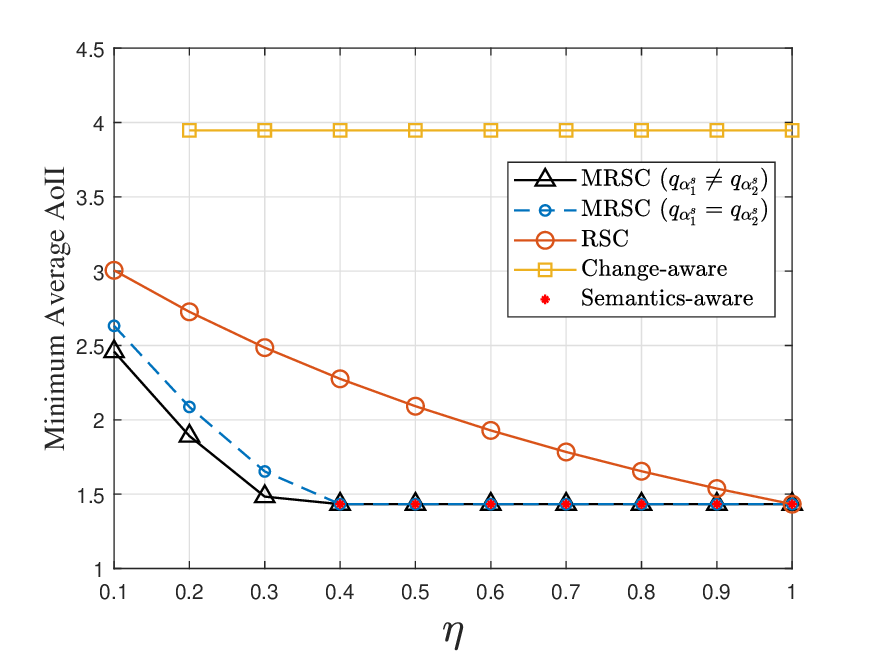}
			\label{MinAvg_AoII_VarMu_p0.2q0.1ps0.1}}
		\subfigure[$p_{{\text{s}}} = 0.9$]{\centering
			\includegraphics[trim=0.5cm 0.05cm 1.1cm 0.7cm,width=0.48\linewidth, clip]{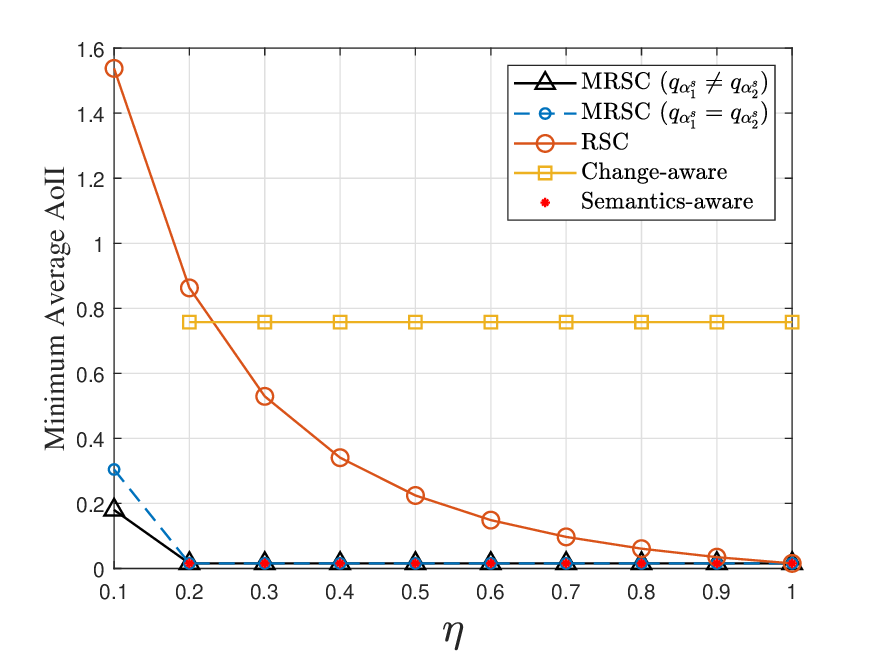}
		}
		\caption{Minimum Average AoII as a function of $\eta$ for $p = 0.2$, and $q = 0.1$.}
		\label{MinAvgAoII_VarMu_p0.2q0.1}
	\end{figure}
	\begin{figure}[!h]
		\centering
		\subfigure[$p_{{\text{s}}} = 0.1$ ]{\includegraphics[trim=0.36cm 0.05cm 1.1cm 0.7cm,width=0.48\linewidth, clip]{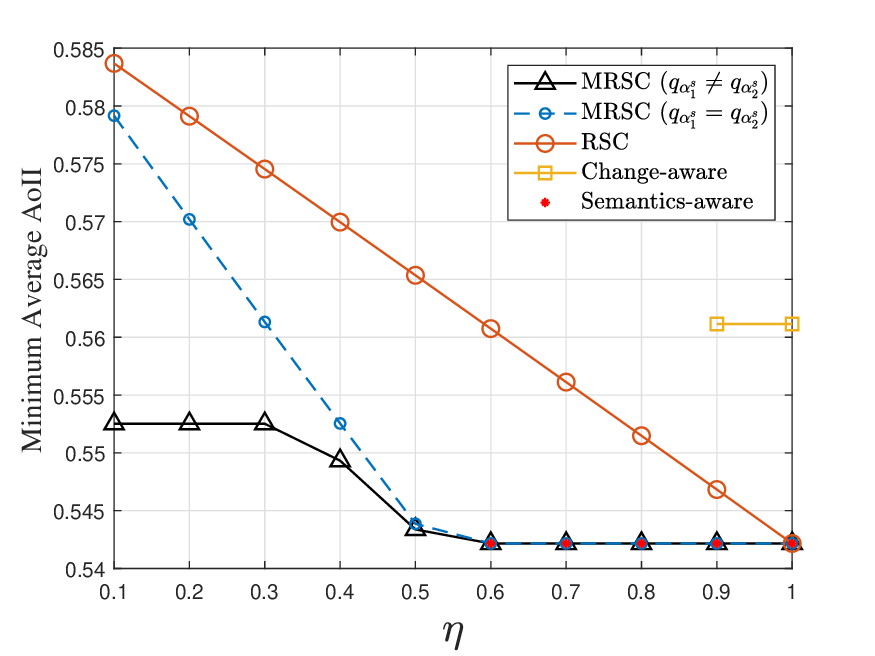}
			\label{MinAvg_AoII_VarMu_p0.8q0.9ps0.1}}
		\subfigure[$p_{{\text{s}}} = 0.9$]{\centering
			\includegraphics[trim=0.36cm 0.05cm 1.1cm 0.7cm,width=0.48\linewidth, clip]{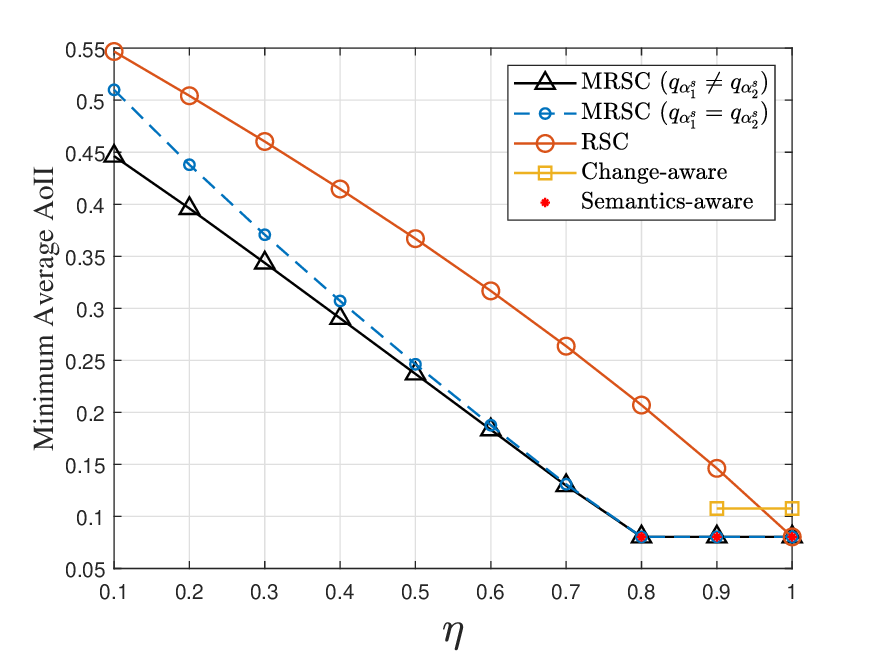}
		}
		\caption{Minimum Average AoII as a function of $\eta$ for $p = 0.8$, and $q = 0.9$.}
		\label{MinAvg_AoII_VarMu_p0.8q0.9}
	\end{figure}
	Figs. \ref{MinAvgAoII_VarMu_p0.2q0.1} and \ref{MinAvg_AoII_VarMu_p0.8q0.9} depict the minimum average AoII as a function of $\eta$ for slowly and rapidly changing sources, considering various values of $p_{s}$. These figures illustrate that the optimal MRSC policy outperforms other policies, particularly when the total time-averaged sampling cost threshold $\eta$ is low and the source exhibits rapid changes. This is because when the source changes rapidly, the semantics-aware and change-aware policies generate more samples than the other policies. As a result, for a smaller value of $\eta$, the time-averaged sampling cost constraint will be violated, and optimal solutions for these policies do not exist. Furthermore, in the optimal MRSC policy, the sampler refrains from sampling when the current state of the source matches the previous state of the reconstructed source. Hence, this policy can generate samples with a higher probability than the optimal RSC policy, where sampling is conducted with probability $\eta$ at each time slot. Consequently, the optimal MRSC policy performs better than the optimal RSC policy.
	\vspace{-0.2cm}
	
	\section{Conclusion}
	We studied a time-slotted communication system where a sampler performs sampling, and the transmitter forwards the samples over an unreliable wireless channel to a receiver that monitors the evolution of a two-state DTMC. We introduced a modified randomized stationary sampling and transmission policy. We then proposed two new metrics, namely VIA and AoIV, and we analyzed the system's average performance regarding these metrics and the AoII for the proposed modified randomized stationary, as well as for randomized stationary, change-aware, and semantics-aware policies. Furthermore, three cost-constrained optimization problems were formulated to find the optimal randomized stationary and modified randomized stationary policies. Our results showed that in terms of the average VIA, the optimal randomized stationary policy outperforms the change-aware policy for slowly and rapidly evolving sources. However, the change-aware policy could perform better for moderately changing sources under certain conditions. Furthermore, in terms of the average AoIV and AoII, the optimal modified randomized stationary policy outperforms all other policies across varying states of the source and channel. This superior performance holds irrespective of whether the source exhibits slow, moderate, or rapid changes and regardless of the channel quality.
	\vspace{-0.4cm}
	\bibliographystyle{IEEEtran}
	\bibliography{ref}

\begin{thebibliography}{10}
\providecommand{\url}[1]{#1}
\csname url@samestyle\endcsname
\providecommand{\newblock}{\relax}
\providecommand{\bibinfo}[2]{#2}
\providecommand{\BIBentrySTDinterwordspacing}{\spaceskip=0pt\relax}
\providecommand{\BIBentryALTinterwordstretchfactor}{4}
\providecommand{\BIBentryALTinterwordspacing}{\spaceskip=\fontdimen2\font plus
\BIBentryALTinterwordstretchfactor\fontdimen3\font minus
  \fontdimen4\font\relax}
\providecommand{\BIBforeignlanguage}[2]{{%
\expandafter\ifx\csname l@#1\endcsname\relax
\typeout{** WARNING: IEEEtran.bst: No hyphenation pattern has been}%
\typeout{** loaded for the language `#1'. Using the pattern for}%
\typeout{** the default language instead.}%
\else
\language=\csname l@#1\endcsname
\fi
#2}}
\providecommand{\BIBdecl}{\relax}
\BIBdecl

\bibitem{abd2019role}
M.~A. Abd-Elmagid, N.~Pappas, and H.~S. Dhillon, ``{On the role of age of
  information in the {Internet of Things}},'' \emph{IEEE Communications
  Magazine}, vol.~57, no.~12, pp. 72--77, 2019.

\bibitem{shreedhar2019age}
T.~Shreedhar, S.~K. Kaul, and R.~D. Yates, ``{An age control transport protocol
  for delivering fresh updates in the {Internet-of-Things}},'' in \emph{IEEE
  20th International Symposium on" A World of Wireless, Mobile and Multimedia
  Networks"(WoWMoM)}, 2019, pp. 1--7.

\bibitem{kountouris2021semantics}
M.~Kountouris and N.~Pappas, ``{Semantics-empowered communication for networked
  intelligent systems},'' \emph{IEEE Commun. Mag.}, 2021.

\bibitem{tolga21SP}
M.~Kalfa, M.~Gok, A.~Atalik, B.~Tegin, T.~M. Duman, and O.~Arikan, ``{Towards
  goal-oriented semantic signal processing: Applications and future
  challenges},'' \emph{Digital Signal Processing}, vol. 119, 2021.

\bibitem{popovski2020semantic}
P.~Popovski, O.~Simeone, F.~Boccardi, D.~G{\"u}nd{\"u}z, and O.~Sahin,
  ``Semantic-effectiveness filtering and control for post-{5G} wireless
  connectivity,'' \emph{Journal of the Indian Institute of Science}, 2020.

\bibitem{PetarProc2022}
P.~Popovski and \textit{et al.}, ``{A Perspective on Time Toward Wireless
  6{G}},'' \emph{Proceedings of the IEEE}, vol. 110, no.~8, 2022.

\bibitem{GunduzJSAC23}
D.~Gündüz, Z.~Qin, I.~E. Aguerri, H.~S. Dhillon, Z.~Yang, A.~Yener, K.~K.
  Wong, and C.-B. Chae, ``{Beyond Transmitting Bits: Context, Semantics, and
  Task-Oriented Communications},'' \emph{IEEE J. Sel. Areas Commun.}, vol.~41,
  no.~1, pp. 5--41, 2023.

\bibitem{kaul2012Real}
S.~Kaul, R.~Yates, and M.~Gruteser, ``{Real-time status: How often should one
  update?}'' in \emph{IEEE Conference on Computer Communications (INFOCOM)},
  2012, pp. 2731--2735.

\bibitem{stamatakis2019control}
G.~Stamatakis, N.~Pappas, and A.~Traganitis, ``{Control of status updates for
  energy harvesting devices that monitor processes with alarms},'' in
  \emph{IEEE Globecom Workshops (GC Wkshps)}, 2019.

\bibitem{pappas2022agebook}
N.~Pappas, M.~A. Abd-Elmagid, B.~Zhou, W.~Saad, and H.~S. Dhillon, \emph{{Age
  of Information: Foundations and Applications}}.\hskip 1em plus 0.5em minus
  0.4em\relax Cambridge University Press, 2023.

\bibitem{stamatakis2024semantics}
G.~Stamatakis, N.~Pappas, A.~Fragkiadakis, N.~Petroulakis, and A.~Traganitis,
  ``{Semantics-Aware Active Fault Detection in Status Updating Systems},''
  \emph{IEEE Open Journal of the Communications Society}, 2024.

\bibitem{yates2021age}
R.~D. Yates, ``{The age of gossip in networks},'' in \emph{IEEE International
  Symposium on Information Theory (ISIT)}, 2021, pp. 2984--2989.

\bibitem{yates2021timely}
------, ``{Timely gossip},'' in \emph{IEEE 22nd International Workshop on
  Signal Processing Advances in Wireless Communications (SPAWC)}, 2021.

\bibitem{buyukatesversion}
B.~Buyukates, M.~Bastopcu, and S.~Ulukus, ``{Version age of information in
  clustered gossip networks},'' \emph{IEEE Journal on Selected Areas in
  Information Theory}, vol.~3, no.~1, pp. 85--97, 2022.

\bibitem{kaswan2022timely}
P.~Kaswan and S.~Ulukus, ``{Timely gossiping with file slicing and network
  coding},'' in \emph{IEEE International Symposium on Information Theory
  (ISIT)}, 2022, pp. 928--933.

\bibitem{kaswan2022susceptibility}
------, ``{Susceptibility of age of gossip to timestomping},'' in \emph{IEEE
  Information Theory Workshop (ITW)}, 2022, pp. 398--403.

\bibitem{kaswan2022age}
------, ``{Age of gossip in ring networks in the presence of jamming
  attacks},'' in \emph{56th Asilomar Conference on Signals, Systems, and
  Computers}, 2022, pp. 1055--1059.

\bibitem{mitra2023age}
P.~Mitra and S.~Ulukus, ``{Age-Aware Gossiping in Network Topologies},''
  \emph{arXiv preprint arXiv:2304.03249}, 2023.

\bibitem{abd2023distribution}
M.~A. Abd-Elmagid and H.~S. Dhillon, ``{Distribution of the Age of Gossip in
  Networks},'' \emph{Entropy}, vol.~25, no.~2, p. 364, 2023.

\bibitem{KaswanTCOM2023}
P.~Kaswan and S.~Ulukus, ``{Timestomping Vulnerability of Age-Sensitive Gossip
  Networks},'' \emph{IEEE Transactions on Communications}, 2024.

\bibitem{KaswanJSAC2023}
------, ``{How Robust are Timely Gossip Networks to Jamming Attacks?}''
  \emph{IEEE Journal on Selected Areas in Information Theory}, 2023.

\bibitem{bastopcu2023role}
M.~Bastopcu, S.~R. Etesami, and T.~Ba{\c{s}}ar, ``The role of gossiping in
  information dissemination over a network of agents,'' \emph{Entropy}, 2023.

\bibitem{mitra2023learning}
P.~Mitra and S.~Ulukus, ``{A Learning Based Scheme for Fair Timeliness in
  Sparse Gossip Networks},'' \emph{arXiv preprint arXiv:2310.01396}, 2023.

\bibitem{delfani2023version}
E.~Delfani and N.~Pappas, ``Version age-optimal cached status updates in a
  gossiping network with energy harvesting sensor,'' in \emph{{21st
  International Symposium on Modeling and Optimization in Mobile, Ad Hoc, and
  Wireless Networks (WiOpt)}}, 2023, pp. 143--150.

\bibitem{karevvanavar2023version}
G.~Karevvanavar, H.~Pable, O.~Patil, R.~Bhat, and N.~Pappas, ``{Version Age of
  Information Minimization over Fading Broadcast Channels},'' \emph{arXiv
  preprint arXiv:2311.09975}, 2023.

\bibitem{maatouk2020age}
A.~Maatouk, S.~Kriouile, M.~Assaad, and A.~Ephremides, ``{The age of incorrect
  information: A new performance metric for status updates},'' \emph{IEEE/ACM
  Transactions on Networking}, 2020.

\bibitem{pappas2021goal}
N.~Pappas and M.~Kountouris, ``{Goal-oriented communication for real-time
  tracking in autonomous systems},'' in \emph{IEEE International Conference on
  Autonomous Systems (ICAS)}, 2021.

\bibitem{jayanth23}
{Jayanth S., N. Pappas, R. Bhat}, ``{Distortion Minimization with Age of
  Information and Cost Constraints},'' \emph{21st International Symposium on
  Modeling and Optimization in Mobile, Ad Hoc, and Wireless Networks (WiOpt)},
  Aug. 2023.

\bibitem{MSalimnejadTCOM2024}
M.~Salimnejad, M.~Kountouris, and N.~Pappas, ``{Real-time Reconstruction of
  Markov Sources and Remote Actuation over Wireless Channels},'' \emph{IEEE
  Trans. Commun.}, 2024.

\bibitem{MSalimnejadJCN2023}
------, ``{State-aware real-time tracking and remote reconstruction of a Markov
  source},'' \emph{Journal of Communications and Networks}, 2023.

\bibitem{fountoulakis2023goal}
E.~Fountoulakis, N.~Pappas, and M.~Kountouris, ``{Goal-oriented policies for
  cost of actuation error minimization in wireless autonomous systems},''
  \emph{IEEE Communications Letters}, 2023.

\bibitem{luo2024semantic}
J.~Luo and N.~Pappas, ``{Semantic-Aware Remote Estimation of Multiple Markov
  Sources Under Constraints},'' \emph{arXiv preprint arXiv:2403.16855}, 2024.

\bibitem{cocco2023remote}
G.~Cocco, A.~Munari, and G.~Liva, ``{Remote Monitoring of Two-State Markov
  Sources via Random Access Channels: An Information Freshness vs. State
  Estimation Entropy Perspective},'' \emph{IEEE Journal on Selected Areas in
  Information Theory}, vol.~4, pp. 651--666, 2023.

\bibitem{nayyar2013optimal}
A.~Nayyar, T.~Ba{\c{s}}ar, D.~Teneketzis, and V.~V. Veeravalli, ``{Optimal
  strategies for communication and remote estimation with an energy harvesting
  sensor},'' \emph{IEEE Trans. Automat. Contr.}, 2013.

\bibitem{chakravorty2014optimal}
J.~Chakravorty and A.~Mahajan, ``{On the optimal thresholds in remote state
  estimation with communication costs},'' in \emph{IEEE Conference on Decision
  and Control (CDC)}, 2014.

\bibitem{shi2012scheduling}
L.~Shi and H.~Zhang, ``{Scheduling two {Gauss--Markov} systems: An optimal
  solution for remote state estimation under bandwidth constraint},''
  \emph{IEEE Transactions on Signal Processing}, 2012.

\bibitem{wu2018optimal}
S.~Wu, X.~Ren, S.~Dey, and L.~Shi, ``{Optimal scheduling of multiple sensors
  over shared channels with packet transmission constraint},''
  \emph{Automatica}, vol.~96, pp. 22--31, 2018.

\bibitem{chakravorty2015distortion}
J.~Chakravorty and A.~Mahajan, ``{Distortion-transmission trade-off in
  real-time transmission of Markov sources},'' in \emph{IEEE Information Theory
  Workshop (ITW)}, 2015.

\bibitem{chakravorty2016fundamental}
------, ``{Fundamental limits of remote estimation of autoregressive Markov
  processes under communication constraints},'' \emph{IEEE Transactions on
  Automatic Control}, vol.~62, no.~3, pp. 1109--1124, 2016.

\bibitem{chakravorty2019remote}
------, ``{Remote estimation over a packet-drop channel with Markovian
  state},'' \emph{IEEE Trans. Automat. Contr.}, 2019.

\end{thebibliography}
	
	\vspace{-0.2cm}
	\appendix
	
	\subsection{Proof of Lemma {\ref{theorem_VAoI}}}
	\label{Appendix_LemmaPij_VAoI}
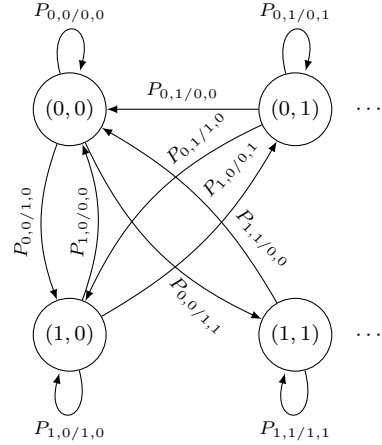
\begin{figure}[!h]
		\centering
		\begin{tikzpicture}[start chain=going left,->,>=latex,on grid,auto]
			\footnotesize
			\footnotesize
			\node[on chain]             at (3,0)(h) {$\cdots$};
			\node[state, on chain]              at (3,0) (2) {$(0,1)$};
			\node[state, on chain]             at (0,0)(1) {$(0,0)$};
			\node[state, on chain]             at (0,-3)(4) {$(1,0)$};
			\node[state, on chain]             at (3,-3)(5) {$(1,1)$};
			\node[on chain]             at (4,-3)(g) {$\cdots$};
			
			\draw[>=latex]
			(1) edge[loop above] node {\scriptsize${P_{0,0/0,0}}$}   (1)
			
			(1) edge [bend right=20, left] node [pos=0.5, left]{\rotatebox{90}{\scriptsize$P_{0,0/1,0}$}}(4)
			(1) edge[bend right=20] node [pos=0.9, left]{\rotatebox{-45}{\scriptsize$P_{0,0/1,1}$}}   (5)
			(5) edge[bend right=15] node [pos=0.1, above]{\rotatebox{-45}{\scriptsize$P_{1,1/0,0}$}}   (1)
			(2) edge[loop above] node {\scriptsize${P_{0,1/0,1}}$}   (2)
			(2) edge[above] node {\scriptsize${P_{0,1/0,0}}$}   (1)
			(2) edge[bend right=20] node [pos=0.1, left]{\rotatebox{45}{\scriptsize$P_{0,1/1,0}$}}   (4)
			(4) edge[bend right=20, right] node [pos=0.5, left]{\rotatebox{90}{\scriptsize$P_{1,0/0,0}$}}   (1)
			(4) edge[loop below] node {\scriptsize${P_{1,0/1,0}}$}   (4)
			(4) edge[bend right=15] node [pos=.95, left,yshift=-2mm]{\rotatebox{50}{\scriptsize$P_{1,0/0,1}$}}   (2)
			(5) edge[loop below] node {\scriptsize${P_{1,1/1,1}}$}   (5)
			;
		\end{tikzpicture}
		\caption{Two-dimensional DTMC describing the joint status of the original source regarding the current state of the VIA using a two-state information source model, i.e., $\big(X(t),{\text{VIA}}(t)\big)$.}
		\label{Ver_AoI}
	\end{figure}

	To obtain $\pi_{j,i}$ we depict the two-dimensional DTMC describing the joint status of the original source regarding the current
	state of the VIA, i.e.,  $\big(X(t), {\text{VIA}}(t)\big)$ in Fig. \ref{Ver_AoI}, where the transition probabilities $P_{i,j/m,n} = \mathrm{Pr}\big[X(t\!+\!1) \!=\! m, {\text{VIA}}(t\!+\!1)\!=\!n\big|X(t) \!=\! i,{\text{VIA}}(t)\!=\!j  \big]$, $\forall i,m\in\{0,1\}$ and $\forall j,n\in\{0,1,\cdots\}$ are given by
	\begin{align}
		\label{trans_prob_VAoI}
		P_{0,0/0,0}\!&=\! 1-p,\hspace{1.1cm}
		P_{0,j/0,0}\!=\! (1\!-\!p)p_{\alpha^{\text{s}}}p_{\text{s}},\notag\\
		P_{0,j/1,0} &=pp_{\alpha^{\text{s}}}p_{\text{s}},\hspace{0.6cm}
		P_{0,j/0,j+1} \!=\!0,\notag\\
		P_{0,j/1,j+1}\!&=p(1\!-\!p_{\alpha^{\text{s}}}p_{\text{s}}),\hspace{0.13cm}
		P_{0,j/0,j}=(1\!-\!p)(1\!-\!p_{\alpha^{\text{s}}}p_{\text{s}}),\notag\\
		P_{1,0/1,0} 
		&= 1-q,\hspace{1.cm}
		P_{1,j/1,0} 
		= (1-q)p_{\alpha^{\text{s}}}p_{\text{s}},\notag\\
		P_{1,j/1,j+1}&=0,\hspace{1.6cm}
		P_{1,j/0,0} =qp_{\alpha^{\text{s}}}p_{\text{s}},\notag\\
		P_{1,j/0,j+1} &=q(1-p_{\alpha^{\text{s}}}p_{\text{s}}),
		P_{1,j/1,j} =(1\!-\!q)(1\!-\!p_{\alpha^{\text{s}}}p_{\text{s}}).
	\end{align}
	Now, using Fig. \ref{Ver_AoI} and \eqref{trans_prob_VAoI}, one can derive the state stationary distribution $\pi_{j,i}$ $\forall j\in\{0,1\}$ and $i\geqslant 0$ as follows
	\begin{align}
		\!\!\pi_{0,i}\!\!&=\!\!\frac{p^kq^wp_{\alpha^{\text{s}}}p_{s}\big(1\!-\!p_{\alpha^{\text{s}}}p_{s}\big)^i}{(p\!+\!q)\big(p\!+\!(1\!-\!p)p_{\alpha^{\text{s}}}p_{s}\big)^w\big(q\!+\!(1\!-\!q)p_{\alpha^{\text{s}}}p_{s}\big)^k},
		i\!=\!0,1,\ldots.\notag\\
		\!\!\pi_{1,i}\!\!&=\!\!\frac{p^wq^kp_{\alpha^{\text{s}}}p_{s}\big(1\!-\!p_{\alpha^{\text{s}}}p_{s}\big)^i}{(p\!+\!q)\big(p\!+\!(1\!-\!p)p_{\alpha^{\text{s}}}p_{s}\big)^k\big(q\!+\!(1\!-\!q)p_{\alpha^{\text{s}}}p_{s}\big)^w}, i\!=\!0,1,\ldots.\label{RandStat_piij_VAoI}
	\end{align}
	where $k$, and $w$ are given by
	\begin{align}
		\label{kw_AoI2}
		k=
		\begin{cases}
			\frac{i}{2}, &\mod\{i,2\}=0,\\
			\frac{i+1}{2}, &\mod\{i,2\}\neq 0.
		\end{cases}\notag\\
		\hspace{0.2cm}w=
		\begin{cases}
			\frac{i+2}{2}, &\mod\{i,2\}=0,\\
			\frac{i+1}{2}, &\mod\{i,2\}\neq 0.
		\end{cases}
	\end{align}
	Similarly, for the change-aware policy $\pi_{0,i}$ and $\pi_{1,i}$ in \eqref{RandStat_piij_VAoI} can be written as
	\begin{align}
		\label{pi_CA_AoI}
		\pi_{0,i} &= \frac{qp_{\text{s}}(1-p_{\text{s}})^i}{p+q}, \hspace{0.2cm} i =0, 1, \cdots.\notag\\
		\pi_{1,i} &= \frac{pp_{\text{s}}(1-p_{\text{s}})^i}{p+q}, \hspace{0.2cm} i =0, 1, \cdots.
	\end{align}
	Using \eqref{pi_CA_AoI}, we can obtain the average VIA, $\overline{\text{VIA}}$, for the change-aware policy as
	\begin{align}
		\label{Avg_AoIV_CA}
		\overline{\text{VIA}} = \sum_{i=0}^{\infty} i\mathrm{Pr}[{\text{VIA}}(t)=i] =  \sum_{i=0}^{\infty} i\big(\pi_{0,i}+\pi_{1,i}\big)=\frac{1-p_{\text{s}}}{p_{\text{s}}}.
	\end{align}
	\vspace{-0.4cm}
	\subsection{Proof of Lemma {\ref{piijk_VAoII_RS}}}
	\label{Appendix_piijk_VAoII_RS}
	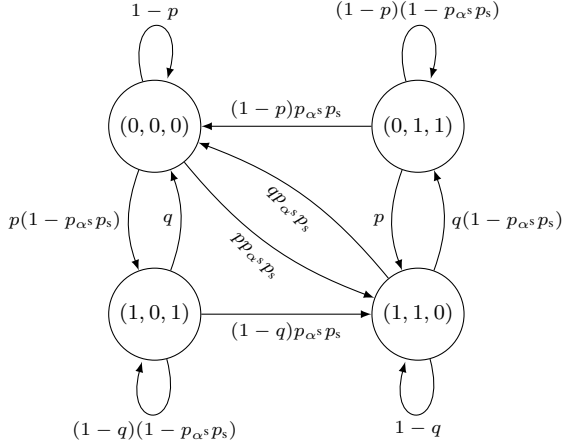
\begin{figure}[!t]
		\centering
		\begin{tikzpicture}[start chain=going left,->,>=latex,on grid,auto]
			\footnotesize
			\footnotesize
			\node[state, on chain]              at (2.5,0) (2) {$(0,1,1)$};
			\node[state, on chain]             at (0,0)(1) {$(0,0,0)$};
			\node[state, on chain]             at (0,-2.5)(3) {$(1,0,1)$};
			\node[state, on chain]             at (3.5,-2.5)(4) {$(1,1,0)$};
			
			\draw[>=latex]
			(1) edge[loop above] node {\scriptsize${1-p}$}   (1)
			
			(1) edge [bend right=20, left] node [pos=0.5, left]{{\scriptsize$p(1-p_{\alpha^{\text{s}}}p_{\text{s}})$}}(3)
			(1) edge[bend right=15] node [pos=0.6, left]{\rotatebox{-45}{\scriptsize$pp_{\alpha^{\text{s}}}p_{\text{s}}$}}   (4)
			
			(2) edge[loop above] node {\scriptsize${(1-p)(1-p_{\alpha^{\text{s}}}p_{\text{s}})}$}   (2)
			(2) edge[above] node {\scriptsize${(1-p)p_{\alpha^{\text{s}}}p_{\text{s}}}$}   (1)
			(2) edge[bend right=20] node [pos=0.5, left]{{\scriptsize$p$}}   (4)
			(3) edge[bend right=20, right] node [pos=0.5, left]{{\scriptsize$q$}}   (1)
			(3) edge[loop below] node {\scriptsize${(1-q)(1-p_{\alpha^{\text{s}}}p_{\text{s}})}$}   (3)
			(3) edge[below] node {\scriptsize${(1-q)p_{\alpha^{\text{s}}}p_{\text{s}}}$}  (4)
			(4) edge[loop below] node {\scriptsize${1-q}$}   (4)
			(4) edge[bend right=20] node [pos=0.5, right]{{\scriptsize$q(1-p_{\alpha^{\text{s}}}p_{\text{s}})$}}   (2)
			(4) edge[bend right=15] node [pos=0.4, left]{\rotatebox{-45}{\scriptsize$qp_{\alpha^{\text{s}}}p_{\text{s}}$}}   (1)
			;
		\end{tikzpicture}
		\caption{Three-dimensional DTMC describing the joint status of the original and reconstructed source regarding the current state of the AoIV, i.e., $\big(X(t),\hat{X}(t),{\text{AoIV}}(t)\big)$.}
		\label{3DMC_VAoII}
	\end{figure}
	\par To derive $\pi_{i,j,k}, \forall 
	i,j,k \in\{0,1\}$, we represent the three-dimensional DTMC describing the joint status of the original and reconstructed source regarding the current state of the AoIV, i.e., $\big(X(t),\hat{X}(t),{\text{AoIV}}(t)\big)$ as Fig. \ref{3DMC_VAoII}. Now, using Fig. \ref{3DMC_VAoII}, we can obtain the state stationary $\pi_{i,j,k}$ for the RS policy as		
	\begin{align}
		\label{VAoII_pi_ijk_RS_proof}
		\pi_{0,0,0} &= \mathrm{Pr}\big[X(t)=0,\hat{X}(t)=0,{\text{AoIV}}(t)=0\big]\notag\\
		&=\frac{q\big[q+(1-q)p_{\alpha^{\text{s}}}p_{s}\big]}{(p+q)\big[p+q+(1-p-q)p_{\alpha^{\text{s}}}p_{s}\big]}, \notag\\
		\pi_{0,1,1} &= \mathrm{Pr}\big[X(t)=0,\hat{X}(t)=1,{\text{AoIV}}(t)=1\big]\notag\\
		&=\frac{pq\big(1-p_{\alpha^{\text{s}}}p_{s}\big)}{(p+q)\big[p+q+(1-p-q)p_{\alpha^{\text{s}}}p_{s}\big]},\notag\\
		\pi_{1,1,0} &= \mathrm{Pr}\big[X(t)=1,\hat{X}(t)=1,{\text{AoIV}}(t)=0\big]\notag\\
		&=\frac{p\big[p+(1-p)p_{\alpha^{\text{s}}}p_{s}\big]}{(p+q)\big[p+q+(1-p-q)p_{\alpha^{\text{s}}}p_{s}\big]},\notag\\
		\pi_{1,0,1} &= \mathrm{Pr}\big[X(t)=1,\hat{X}(t)=0,{\text{AoIV}}(t)=1\big]\notag\\
		&=\frac{pq\big(1-p_{\alpha^{\text{s}}}p_{s}\big)}{(p+q)\big[p+q+(1-p-q)p_{\alpha^{\text{s}}}p_{s}\big]},\notag\\
		\pi_{0,0,1}&=\pi_{0,1,0}=\pi_{1,0,0}=\pi_{1,1,1}=0.
	\end{align}
	Similarly, for the MRS policy, we can write \eqref{VAoII_pi_ijk_RS_proof} as follows
	\begin{align}
		\label{VAoII_pi_ijk_MRS_Proof}
		\pi_{0,0,0}
		&=\frac{q\big[q_{\alpha^{\text{s}}_{2}}+p\big(q_{\alpha^{\text{s}}_{1}}-q_{\alpha^{\text{s}}_{2}}\big)\big]\big[q+(1-q)q_{\alpha^{\text{s}}_{2}}p_{s}\big]}{(p+q)F\big(q_{\alpha^{\text{s}}_{1}},q_{\alpha^{\text{s}}_{2}}\big)},\notag\\
		\pi_{0,1,1}
		&=\frac{pq\big(1-q_{\alpha^{\text{s}}_{1}}p_{s}\big)\big[q_{\alpha^{\text{s}}_{2}}+q\big(q_{\alpha^{\text{s}}_{1}}-q_{\alpha^{\text{s}}_{2}}\big)\big]}{(p+q)F\big(q_{\alpha^{\text{s}}_{1}},q_{\alpha^{\text{s}}_{2}}\big)},\notag\\
		\pi_{1,1,0}
		&=\frac{p\big[q_{\alpha^{\text{s}}_{2}}+q\big(q_{\alpha^{\text{s}}_{1}}-q_{\alpha^{\text{s}}_{2}}\big)\big]\big[p+(1-p)q_{\alpha^{\text{s}}_{2}}p_{s}\big]}{(p+q)F\big(q_{\alpha^{\text{s}}_{1}},q_{\alpha^{\text{s}}_{2}}\big)},\notag\\
		\pi_{1,0,1}
		&=\frac{pq\big(1-q_{\alpha^{\text{s}}_{1}}p_{s}\big)\big[q_{\alpha^{\text{s}}_{2}}+p\big(q_{\alpha^{\text{s}}_{1}}-q_{\alpha^{\text{s}}_{2}}\big)\big]}{(p+q)F\big(q_{\alpha^{\text{s}}_{1}},q_{\alpha^{\text{s}}_{2}}\big)},\notag\\
		\pi_{0,0,1}&=\pi_{0,1,0}=\pi_{1,0,0}=\pi_{1,1,1}=0.
	\end{align}
	where $F(\cdot,\cdot)$ in \eqref{VAoII_pi_ijk_MRS_Proof} is given by
	\begin{align}
		\label{F_MRS_Proof}
		F\big(q_{\alpha^{\text{s}}_{1}},q_{\alpha^{\text{s}}_{2}}\big) &= (1-p)(1-q)p_{s}q^{2}_{\alpha^{\text{s}}_{2}}+(p+q-2pq)q_{\alpha^{\text{s}}_{2}}\notag\\
		&+pq(2-p_{s}q_{\alpha^{\text{s}}_{1}})q_{\alpha^{\text{s}}_{1}}.
	\end{align}
	Furthermore, for the change-aware policy,  one can write \eqref{VAoII_pi_ijk_RS_proof} as follows
	\begin{subequations}
		\label{VAoII_pi_ijk_CA_proof}
		\begin{align}
			\pi_{0,0,0} &= \frac{q}{(p+q)(2-p_{\text{s}})},
			\pi_{0,1,1} = \frac{q(1-p_{\text{s}})}{(p+q)(2-p_{\text{s}})},\label{pi011_CA}\\
			\pi_{1,1,0} &= \frac{p}{(p+q)(2-p_{\text{s}})},
			\pi_{1,0,1} = \frac{p(1-p_{\text{s}})}{(p+q)(2-p_{\text{s}})}\label{pi101_CA},\\
			\pi_{0,0,1}&=\pi_{0,1,0}=\pi_{1,0,0}=\pi_{1,1,1}=0.
		\end{align}
	\end{subequations}
	Now, using \eqref{VAoII_pi_ijk_CA_proof}, the average AoIV for the change-aware policy is given by
	\begin{align}
		\label{Avg_AoIVI_CA}
		\overline{\text{AoIV}} &= \sum_{i=1}^{\infty} i\mathrm{Pr}[{\text{AoIV}}(t)=i]=\pi_{0,1,1}+\pi_{1,0,1}=  \frac{1-p_{\text{s}}}{2-p_{\text{s}}}.
	\end{align}
	Moreover, for the semantics-aware policy, we can obtain \eqref{VAoII_pi_ijk_CA_proof} as follows
	\begin{subequations}
		\label{VAoII_pi_ijk_SA_proof}
		\begin{align}
			\pi_{0,0,0} &=
			\frac{q\big[q+(1-q)p_{s}\big]}{(p+q)\big[p+q+(1-p-q)p_{s}\big]},\label{pi000_SA}\\
			\pi_{0,1,1} &= \frac{pq\big[1-p_{s}\big]}{(p+q)\big[p+q+(1-p-q)p_{s}\big]},\label{pi011_SA}\\
			\pi_{1,1,0} 
			&=\frac{p\big[p+(1-p)p_{s}\big]}{(p+q)\big[p+q+(1-p-q)p_{s}\big]},\label{pi110_SA}\\
			\pi_{1,0,1} &= \frac{pq\big[1-p_{s}\big]}{(p+q)\big[p+q+(1-p-q)p_{s}\big]}\label{pi101_SA},\\
			\pi_{0,0,1}&=\pi_{0,1,0}=\pi_{1,0,0}=\pi_{1,1,1}=0.
		\end{align}
	\end{subequations}
	Using \eqref{VAoII_pi_ijk_SA_proof}, the average AoIV for the semantics-aware policy can be written as
	\begin{align}
		\label{Avg_AoIVI_SA}
		\overline{\text{AoIV}}&= 
		\sum_{i=1}^{\infty} i\mathrm{Pr}[{\text{AoIV}}(t)=i]=\pi_{0,1,1}+\pi_{1,0,1}\notag\\
		&=\frac{2pq\big(1-p_{s}\big)}{(p+q)\big[p+q+(1-p-\!q)p_{s}\big]}.
	\end{align}
	
	\subsection{Proof of Lemma {\ref{AoIIDistribution_RS}} and {\ref{AoIIDistribution_MRS}} }
	\label{Appendix_AoIIDistribution_RS}
	\par
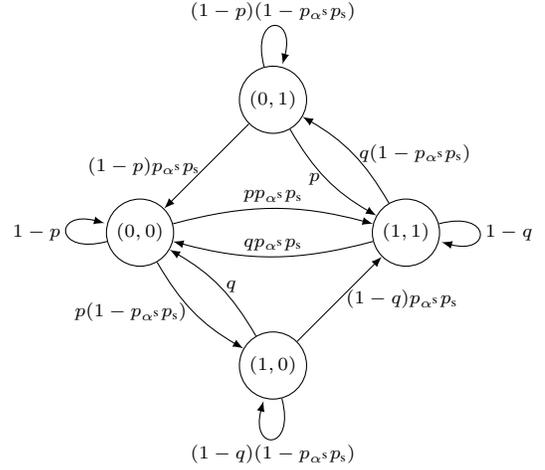
\begin{figure}[!t]
		\centering
		\scriptsize
		\begin{tikzpicture}[start chain=going left,->,>=latex,node distance=2.5cm]
			\node[state]    (A)                     {$(0,0)$};
			\node[state]    (B)[above right of=A]   {$(0,1)$};
			\node[state]    (C)[below right of=A]   {$(1,0)$};
			\node[state]    (D)[below right of=B]   {$(1,1)$};
			\path
			(A) edge[loop left]     node{$1-p$}  (A)
			edge[bend left=15,above]  node{$pp_{\alpha^{\text{s}}}p_{\text{s}}$}  (D)
			edge[bend right=15,left]    node{$p(1-p_{\alpha^{\text{s}}}p_{\text{s}})$}  (C)
			(B) edge[loop above]  node{$(1-p)(1-p_{\alpha^{\text{s}}}p_{\text{s}})$}     (B)
			edge[left=15] node{$(1-p)p_{\alpha^{\text{s}}}p_{\text{s}}$}    (A)
			edge[bend right=15,left] node{$p$}    (D)
			(C) edge[loop below]  node{$(1-q)(1-p_{\alpha^{\text{s}}}p_{\text{s}})$}     (C)
			edge[bend right=15,right] node{$q$}    (A)
			edge[ left=15,right] node{$(1-q)p_{\alpha^{\text{s}}}p_{\text{s}}$}    (D)
			(D) edge[loop right]    node{$1-q$}     (D)
			edge[bend right=15,right]  node{$q(1-p_{\alpha^{\text{s}}}p_{\text{s}})$}  (B)
			edge[bend left=15,above]     node{$qp_{\alpha^{\text{s}}}p_{\text{s}}$}         (A);
		\end{tikzpicture}
		\vspace*{1ex}
		\caption{Two-dimensional DTMC describing the joint status of the original source regarding the current state at the reconstructed source using a two-state information source model.}
		\label{2DimMarkovChain}
	\end{figure}

 At time slot $t$, $\text{AoII} = 0$ denotes the sync state, therefore $\mathrm{Pr}\big[{\text{AoII}}(t)=0\big]$ can be written as
	\begin{align}
		\label{AoII_Sync}
		\mathrm{Pr}\big[{\text{AoII}}(t)=0\big] &= \mathrm{Pr}\big[X(t)=0, \hat{X}(t)=0\big]\notag\\
		&+\mathrm{Pr}\big[X(t)=1, \hat{X}(t)=1\big] = \pi^{\text{A}}_{0,0}+\pi^{\text{A}}_{1,1}.
	\end{align}
	Furthermore,  $\text{AoII}(t)=i\geqslant 1$ indicates that the system was in a sync state at time slot $t-i$, and it has been in an erroneous state from time slots $t-i+1$ to $t$. Therefore, we can calculate $\mathrm{Pr}\big[\text{AoII}(t)=i\big]$ for the RS policy as follows
	\begin{align}
		\label{AoII_NoSync}
		&\mathrm{Pr}\big[{\text{AoII}}(t)\!=\!i\big]\notag\\
		&= \mathrm{Pr}\Big[X(t)\!\neq\! \hat{X}(t), X(t\!-\!1)\!\neq\! \hat{X}(t\!-\!1), \cdots\notag\\
		&, X(t\!-\!i\!+\!1)\neq \hat{X}(t\!-\!i\!+\!1),X(t\!-\!i)= \hat{X}(t\!-\!i)\Big]\notag\\
		&=\mathrm{Pr}\Big[X(t)\!\neq\! \hat{X}(t), X(t\!-\!1)\!\neq\! \hat{X}(t\!-\!1), \cdots\notag\\
		&, X(t\!-\!i\!+\!1)\neq \hat{X}(t\!-\!i\!+\!1)\Big|X(t\!-\!i)= 0,\hat{X}(t\!-\!i)=0\Big]\notag\\
		&\times\mathrm{Pr}\Big[X(t\!-\!i)= 0,\hat{X}(t\!-\!i)=0\Big]+\mathrm{Pr}\Big[X(t)\!\neq\! \hat{X}(t),\cdots\notag\\
		&, X(t\!-\!i\!+\!1)\neq \hat{X}(t\!-\!i\!+\!1)\Big|X(t\!-\!i)= 1,\hat{X}(t\!-\!i)=1\Big]\notag\\
		&\times\mathrm{Pr}\Big[X(t\!-\!i)= 1,\hat{X}(t\!-\!i)=1\Big]\notag\\
		&=\!\!p(1-q)^{i-1}(1\!-\!p_{\alpha^{\text{s}}}p_{\text{s}})^{i}\pi^{\text{A}}_{0,0}\!+\!q(1-p)^{i-1}(1\!-\!p_{\alpha^{\text{s}}}p_{\text{s}})^{i}\pi^{\text{A}}_{1,1},
	\end{align}
	where $\pi^{\text{A}}_{i,j}, \forall i,j\in\{0,1\}$ are the probabilities derived from the stationary distribution of the two-dimensional DTMC describing the joint status of the original source regarding the current state at the reconstructed source, i.e., $\big(X(t), \hat{X}(t)\big)$. Now, using the two-dimensional DTMC depicted in Fig. \ref{2DimMarkovChain}, one can obtain $\pi^{\text{A}}_{i,j}$ as follows
	\begin{align}
		\label{pij_2D_RS}
		\pi^{\text{A}}_{0,0} &=\frac{q\big[q+(1-q)p_{\alpha^{\text{s}}}p_{\text{s}}\big]}{(p+q)\big[p+q+(1-p-q)p_{\alpha^{\text{s}}}p_{\text{s}}\big]},\notag\\\hspace{0.1cm}
		\pi^{\text{A}}_{0,1} &=\frac{pq(1-p_{\alpha^{\text{s}}}p_{\text{s}})}{(p+q)\big[p+q+(1-p-q)p_{\alpha^{\text{s}}}p_{\text{s}}\big]},\notag\\
		\pi^{\text{A}}_{1,0} &=\frac{pq(1-p_{\alpha^{\text{s}}}p_{\text{s}})}{(p+q)\big[p+q+(1-p-q)p_{\alpha^{\text{s}}}p_{\text{s}}\big]},\notag\\
		\pi^{\text{A}}_{1,1} &=\frac{p\big[p+(1-p)p_{\alpha^{\text{s}}}p_{\text{s}}\big]}{(p+q)\big[p+q+(1-p-q)p_{\alpha^{\text{s}}}p_{\text{s}}\big]}.
	\end{align}
	Using \eqref{pij_2D_RS}, $\mathrm{Pr}\big[\text{AoII}(t)=i\big]$ given in \eqref{AoII_Sync} and \eqref{AoII_NoSync} can be written as
	\begin{align}
		\label{PrAoII_RS}
		&\mathrm{Pr}\big[\text{AoII}(t)=i\big] \notag\\
		&\!\!\!=\!\!\! 
		\begin{cases}
			\frac{p^2+q^2+(p+q-p^2-q^2)p_{\alpha^{\text{s}}}p_{\text{s}}}{(p+q)\big[p+q+(1-p-q)p_{\alpha^{\text{s}}}p_{\text{s}}\big]},&\!\!\!\!i\!=\!0,\\
			\frac{pq\big(1-p_{\alpha^{\text{s}}}p_{\text{s}}\big)^{i}\big[(1-q)^{i-1}\Phi(q)+(1-p)^{i-1}\Phi(p)\big]}{(p+q)\big[p+q+(1-p-q)p_{\alpha^{\text{s}}}p_{\text{s}}\big]}, &\!\!\!\!i\!\geqslant\! 1.
		\end{cases}
	\end{align}
	where $\Phi(x)=x+(1-x)p_{\alpha^{\text{s}}}p_{\text{s}}$.
	Similarly, for the MRS policy, we can calculate $\pi^{\text{A}}_{i,j}$ as follows
	\begin{align}
		\label{VAoII_pi_ij_MRS_Proof}
		\pi^{\text{A}}_{0,0}
		&=\frac{q\big[q_{\alpha^{\text{s}}_{2}}+p\big(q_{\alpha^{\text{s}}_{1}}-q_{\alpha^{\text{s}}_{2}}\big)\big]\big[q+(1-q)q_{\alpha^{\text{s}}_{2}}p_{s}\big]}{(p+q)F\big(q_{\alpha^{\text{s}}_{1}},q_{\alpha^{\text{s}}_{2}}\big)},\notag\\
		\pi^{\text{A}}_{0,1}
		&=\frac{pq\big(1-q_{\alpha^{\text{s}}_{1}}p_{s}\big)\big[q_{\alpha^{\text{s}}_{2}}+q\big(q_{\alpha^{\text{s}}_{1}}-q_{\alpha^{\text{s}}_{2}}\big)\big]}{(p+q)F\big(q_{\alpha^{\text{s}}_{1}},q_{\alpha^{\text{s}}_{2}}\big)},\notag\\
		\pi^{\text{A}}_{1,0}
		&=\frac{pq\big(1-q_{\alpha^{\text{s}}_{1}}p_{s}\big)\big[q_{\alpha^{\text{s}}_{2}}+p\big(q_{\alpha^{\text{s}}_{1}}-q_{\alpha^{\text{s}}_{2}}\big)\big]}{(p+q)F\big(q_{\alpha^{\text{s}}_{1}},q_{\alpha^{\text{s}}_{2}}\big)}\notag\\
		\pi^{\text{A}}_{1,1}
		&=\frac{p\big[q_{\alpha^{\text{s}}_{2}}+q\big(q_{\alpha^{\text{s}}_{1}}-q_{\alpha^{\text{s}}_{2}}\big)\big]\big[p+(1-p)q_{\alpha^{\text{s}}_{2}}p_{s}\big]}{(p+q)F\big(q_{\alpha^{\text{s}}_{1}},q_{\alpha^{\text{s}}_{2}}\big)},
	\end{align}
	where $F(\cdot,\cdot)$ given in \eqref{VAoII_pi_ij_MRS_Proof} is obtained in \eqref{F_MRS_Proof}.
	Now, using \eqref{VAoII_pi_ij_MRS_Proof}, $\mathrm{Pr}\big[\text{AoII}(t)=0\big]$ is given by
	\begin{align}
		\label{AoII_i0_MRS}
		\mathrm{Pr}\big[\text{AoII}(t)=0\big] = \frac{G\big(q_{\alpha^{\text{s}}_{1}},q_{\alpha^{\text{s}}_{2}}\big)}{(p+q)F\big(q_{\alpha^{\text{s}}_{1}},q_{\alpha^{\text{s}}_{2}}\big)},
	\end{align}
	where $G(\cdot,\cdot)$ in \eqref{AoII_i0_MRS} is given by
	\begin{align}
		\label{G_MRS}
		G\big(q_{\alpha^{\text{s}}_{1}},q_{\alpha^{\text{s}}_{2}}\big) &= q\left[q_{\alpha^{\text{s}}_{2}}+p\left(q_{\alpha^{\text{s}}_{1}}-q_{\alpha^{\text{s}}_{2}}\right)\right]\big[q+(1-q)q_{\alpha^{\text{s}}_{2}}p_{s}\big]\notag\\
		&+p\big[q_{\alpha^{\text{s}}_{2}}+q\big(q_{\alpha^{\text{s}}_{1}}-q_{\alpha^{\text{s}}_{2}}\big)\big]\big[p+(1-p)q_{\alpha^{\text{s}}_{2}}p_{s}\big].
	\end{align}
	Furthermore, using \eqref{VAoII_pi_ij_MRS_Proof} and following the procedure given in \eqref{AoII_NoSync}, $\mathrm{Pr}\big[\text{AoII}(t)=i\big], \forall i\geqslant 1$ for the MRS policy is calculated as
	\begin{align}
		\label{AoII_i_MRS}
		\mathrm{Pr}\big[\text{AoII}(t)=i\big] = \frac{H\big(q_{\alpha^{\text{s}}_{1}},q_{\alpha^{\text{s}}_{2}}\big)}{(p+q)F\big(q_{\alpha^{\text{s}}_{1}},q_{\alpha^{\text{s}}_{2}}\big)},
	\end{align}
	where $H(\cdot,\cdot)$ in \eqref{AoII_i_MRS} is given by
	\begin{align}
		\label{H_MRS}
		&H\big(q_{\alpha^{\text{s}}_{1}},q_{\alpha^{\text{s}}_{2}}\big)\notag\\
		&= pq\big(1-q_{\alpha^{\text{s}}_{1}}p_{s}\big)\big(1-q_{\alpha^{\text{s}}_{2}}p_{s}\big)^{i-1}\bigg(\Big[q_{\alpha^{\text{s}}_{2}}+p\big(q_{\alpha^{\text{s}}_{1}}-q_{\alpha^{\text{s}}_{2}}\big)\Big]\notag\\
		&\times\Big[q+(1-q)q_{\alpha^{\text{s}}_{2}}p_{s}\Big](1-q)^{i-1}+\Big[q_{\alpha^{\text{s}}_{2}}+q\big(q_{\alpha^{\text{s}}_{1}}-q_{\alpha^{\text{s}}_{2}}\big)\Big]\notag\\
		&\times\Big[p+(1-p)q_{\alpha^{\text{s}}_{2}}p_{s}\Big](1-p)^{i-1}\bigg).
	\end{align}
	Now, using \eqref{AoII_i_MRS}, one can obtain the average AoII for the MRS policy as follows
	\vspace{-0.4cm}
	\begin{align}
		\label{AvgAoII_MRS_Proof}
		\!\!\!\!\overline{\text{AoII}} \!\!&=\!\!\sum_{i=1}^{\infty} i\mathrm{Pr}\big[\text{AoII}(t)\!=\!i\big]\notag\\
		&=\!\!\frac{K\big(q_{\alpha^{\text{s}}_{1}},q_{\alpha^{\text{s}}_{2}}\big)}{(p+q)\!\big(q\!+\!(1-q)q_{\alpha^{\text{s}}_{2}}p_{s}\big)\!\big(p\!+\!(1-p)q_{\alpha^{\text{s}}_{2}}p_{s}\big)F\big(q_{\alpha^{\text{s}}_{1}},q_{\alpha^{\text{s}}_{2}}\big)},
	\end{align}
	where $F(\cdot,\cdot)$ given in \eqref{AvgAoII_MRS_Proof} is obtained in \eqref{F_MRS_Proof} and $K(\cdot,\cdot)$ is given by
	\begin{align}
		\label{K_MRS_Proof}
		\!\!\!K\big(q_{\alpha^{\text{s}}_{1}},q_{\alpha^{\text{s}}_{2}}\big) \!\!&=\! pq\big(1-q_{\alpha^{\text{s}}_{1}}p_{s}\big)\bigg[p^{2}\big(q_{\alpha^{\text{s}}_{1}}-q_{\alpha^{\text{s}}_{2}}\big)\big(1-q_{\alpha^{\text{s}}_{2}}p_{s}\big)\notag\\
		\!&+\!pq_{\alpha^{\text{s}}_{2}}\big(1+q_{\alpha^{\text{s}}_{1}}p_{s}-2q_{\alpha^{\text{s}}_{2}}p_{s}\big)+q^2q_{\alpha^{\text{s}}_{1}}\notag\\
		\!&+\!q(1-q)q_{\alpha^{\text{s}}_{2}}\big(1+q_{\alpha^{\text{s}}_{1}}p_{s}\big)\!+\!\big(2-2q+q^{2}\big)p_{s}q^{2}_{\alpha^{\text{s}}_{2}}\bigg].
	\end{align}
	Moreover, for the change-aware policy, $\mathrm{Pr}\big[\text{AoII}(t)=i\big]$ in \eqref{PrAoII_RS} is given by
	\begin{align}
		\label{PrAoII_CA}
		\!\!\!\mathrm{Pr}\big[\text{AoII}(t)=i\big] \!=\! 
		\begin{cases}
			\frac{1}{2-p_{\text{s}}},\hspace{0.2cm} &i=0,\\
			\frac{pq(1-p_{\text{s}})\big[(1-p)(1-q)\big]^{i-1}}{(p+q)(2-p_{\text{s}})},\hspace{0.2cm} &i\geqslant 1.
		\end{cases}
	\end{align}
	Using \eqref{PrAoII_CA}, the average AoII for the change-aware policy is
	\begin{align}
		\label{Avg_AoII_CA}
		\overline{\text{AoII}} &=\sum_{i=1}^{\infty} i\mathrm{Pr}\big[\text{AoII}(t)\!=\!i\big]=\frac{(p^2+q^2)(1-p_{\text{s}})}{pq(p+q)(2-p_{\text{s}})}.
	\end{align}
	Furthermore, for the semantics-aware policy, we can obtain $\mathrm{Pr}\big[\text{AoII}(t)=i\big]$ as follows
	\begin{align}
		\label{PrAoII_SA}
		&\mathrm{Pr}\big[\text{AoII}(t)=i\big] \notag\\
		&= 
		\begin{cases}
			\frac{p^2+q^2+(p+q-p^2-q^2)p_{\text{s}}}{(p+q)\big[p+q+(1-p-q)p_{\text{s}}\big]},\hspace{0.2cm} &i=0,\\
			\frac{pq(1-p_{\text{s}})^{i}\big[(1-q)^{i-1}\Psi(q)+(1-p)^{i-1}\Psi(p)\big]}{(p+q)\big[p+q+(1-p-q)p_{\text{s}}\big]},\hspace{0.2cm} &i\geqslant 1.
		\end{cases}
	\end{align}
	where $\Psi(x)=x+(1-x)p_{s}$.
	Now, using \eqref{PrAoII_SA}, one can calculate the average AoII for the semantics-aware policy as
	\begin{align}
		\label{Avg_AoII_SA}
		\overline{\text{AoII}} &=\sum_{i=1}^{\infty} i\mathrm{Pr}\big[\text{AoII}(t)\!=\!i\big]\notag\\ &=\frac{pq(1-p_{s})\big[p+q+(2-p-q)p_{s}\big]}{(p+q)\Psi(p)\Psi(q)\big[p+q+(1-p-q)p_{s}\big]}.
	\end{align}
	\subsection{Proof of Lemma {\ref{OptimalMRS}}}
	\label{Appendix_MRS_ASamplingCost}
	\par For the MRSC policy, the constraint given in \eqref{Optimization_prob3_MRS_constraint} can be written as
	\begin{align}
		\label{MRS_Cost1}
		&\lim_{T \to \infty}\frac{1}{T}\sum_{t=1}^{T}\delta \mathbbm{1}\{\alpha^{\text{s}}(t)=1\} \notag\\
		&=\! \delta\mathrm{Pr}\Big[\alpha^{\text{s}}(t)\!=\!1\Big| X(t)\!\neq\! \hat{X}(t-1), X(t-1)=\hat{X}(t-1)\Big]\notag\\
		&\times\mathrm{Pr}\Big[X(t)\!\neq\! \hat{X}(t-1),X(t-1)=\hat{X}(t-1)\Big] \notag\\
		&+\! \delta\mathrm{Pr}\Big[\alpha^{\text{s}}(t)\!=\!1\Big| X(t)\!\neq\! \hat{X}(t-1), X(t-1)\neq\hat{X}(t-1)\Big]\notag\\
		&\times\mathrm{Pr}\Big[X(t)\!\neq\! \hat{X}(t-1),X(t-1)\neq\hat{X}(t-1)\Big]\notag\\
		&= \delta q_{\alpha^{\text{s}}_{1}}\mathrm{Pr}\Big[X(t)\!\neq\! \hat{X}(t-1),X(t-1)=\hat{X}(t-1)\Big]\notag\\
		&+\delta q_{\alpha^{\text{s}}_{2}}\mathrm{Pr}\Big[X(t)\!\neq\! \hat{X}(t-1),X(t-1)\neq\hat{X}(t-1)\Big],
	\end{align}
	using the total probability theorem we have
	\begin{align}
		\label{MRS_Cost2}
		&\lim_{T \to \infty}\frac{1}{T}\sum_{t=1}^{T}\delta \mathbbm{1}\{\alpha^{\text{s}}(t)=1\} \notag\\
		&=\delta q_{\alpha^{\text{s}}_{1}}\mathrm{Pr}\Big[X(t)\!\neq\! \hat{X}(t-1)\Big|X(t-1)=\hat{X}(t-1)\Big]\notag\\
		&\times \mathrm{Pr}\Big[X(t-1)=\hat{X}(t-1)\Big]\notag\\
		&+\delta q_{\alpha^{\text{s}}_{2}}\mathrm{Pr}\Big[X(t)\!\neq\! \hat{X}(t-1)\Big|X(t-1)\neq\hat{X}(t-1)\Big]\notag\\
		&\times \mathrm{Pr}\Big[X(t-1)\neq\hat{X}(t-1)\Big]\notag\\
		&=\delta q_{\alpha^{\text{s}}_{1}}\Big(p\pi^{\text{A}}_{0,0}+q\pi^{\text{A}}_{1,1}\Big)+\delta q_{\alpha^{\text{s}}_{2}}\Big((1-p)\pi^{\text{A}}_{0,1}+(1-q)\pi^{\text{A}}_{1,0}\Big)
	\end{align}
	Now, using \eqref{VAoII_pi_ij_MRS_Proof}, \eqref{MRS_Cost2} is given by
	\begin{align}
		\label{MRS_Cost3}
		&\lim_{T \to \infty}\frac{1}{T}\sum_{t=1}^{T}\delta \mathbbm{1}\{\alpha^{\text{s}}(t)=1\} \notag\\
		&=\frac{2pq\delta\Big[q_{\alpha^{\text{s}}_{2}}+p\big(q_{\alpha^{\text{s}}_{1}}-q_{\alpha^{\text{s}}_{2}}\big)\Big]\Big[q_{\alpha^{\text{s}}_{2}}+q\big(q_{\alpha^{\text{s}}_{1}}-q_{\alpha^{\text{s}}_{2}}\big)\Big]}{(p+q)F\big(q_{\alpha^{\text{s}}_{1}},q_{\alpha^{\text{s}}_{2}}\big)},
	\end{align}
	where $F(\cdot,\cdot)$ given in \eqref{MRS_Cost3} is obtained in \eqref{F_MRS_Proof}. Now, assuming $q_{\alpha^{\text{s}}_{1}}=q_{\alpha^{\text{s}}_{2}} = q_{\alpha^{\text{s}}}$, we can simplify \eqref{MRS_Cost3} as follows
	\begin{align}
		\label{MRS_Cost4}
		\!\!\lim_{T \to \infty}\frac{1}{T}\sum_{t=1}^{T}\delta \mathbbm{1}\{\alpha^{\text{s}}(t)=1\}\!=\! \frac{2pq\delta q_{\alpha^{\text{s}}}}{(p+q)\big[p\!+\!q\!+\!(1\!-\!p\!-\!q)q_{\alpha^{\text{s}}}p_{s}\big]}.
	\end{align}
	Using \eqref{MRS_Cost4}, we can write the constraint of the optimization problem given in \eqref{Optimization_prob3_MRS_constraint} as follows
	\begin{align}
		\label{MRS_Cost5}
		\big[2pq-\eta(p+q)(1-p-q)p_{s}\big]q_{\alpha^{\text{s}}}\leqslant\eta(p+q)^{2},
	\end{align}
	where $\eta=\delta_{\text{max}}/\delta$. Now, using \eqref{MRS_Cost5}, and considering that $0\leqslant q_{\alpha^{\text{s}}} \leqslant 1$, the optimal value of the sampling probability for the MRSC policy is given by
	\begin{align}
		\label{pa2_optimal_proof}
		&q^{*}_{\alpha^{\text{s}}}\notag\\
		&\!=\! 
		\begin{cases}
			1,&\!\!\!\!\eta(p\!+\!q)\!(1\!-\!p\!-\!q)p_{s}\!\!\geqslant\!\! 2pq,\notag\\
			\!\text{min}\left\{\!1,\frac{\eta(p+q)^{2}}{2pq-\eta(p+q)\!(1-p-q)p_{s}}\!\right\}\!,&\!\!\!\!\eta(p\!+\!q)\!(1\!-\!p\!-\!q)p_{s}\!\!<\!\! 2pq.
		\end{cases}
	\end{align}
	Furthermore, using (90) and (91) in \cite{MSalimnejadTCOM2024}, the constraint given in \eqref{Optimization_prob3_MRS_constraint} for the change-aware and semantics-aware policies is simplified as $\frac{2pq\delta}{p+q}$ and $\frac{2pq\delta}{(p+q)\big[p+q+(1-p-q)p_{s}\big]}$, respectively.
\end{document}